\documentclass[
    aps,
    pra,
    reprint,
    superscriptaddress,
    longbibliography
]{revtex4-2}

\usepackage{amsmath}
\usepackage{amssymb}
\usepackage{bm}
\usepackage{graphicx}
\usepackage{braket}
\usepackage{xcolor}
\usepackage[colorlinks=true,
            linkcolor=blue,
            citecolor=blue,
            urlcolor=blue]{hyperref}

\begin{document}
% =============================================================================

% -----------------------------------------------------------------------------
% Front matter
% -----------------------------------------------------------------------------

\title{Quantum Thermal Transistor via Nonlinear Environmental Coupling}

\author{Laiba Hayat}
\affiliation{Faculty of Basic Sciences, Ghulam Ishaq Khan Institute of Engineering Sciences and Technology, Topi 23640, District Swabi, Khyber Pakhtunkhwa, Pakistan}
\author{M. Tahir Naseem}
\email{mnaseem16@ku.edu.tr}
\affiliation{Faculty of Basic Sciences, Ghulam Ishaq Khan Institute of Engineering Sciences and Technology, Topi 23640, District Swabi, Khyber Pakhtunkhwa, Pakistan}
\affiliation{National Center for Quantum Computing, University of Engineering and Technology Narowal, 10 km Muridke Road, Near Adda Siraj, Narowal, Pakistan}

\begin{abstract}
Thermal amplification enables sensitive control of heat currents, providing a basis for heat management and thermal signal processing
in quantum devices. For a harmonic oscillator coupled to emitter, collector, and base thermal reservoirs, we recover the absence of 
thermal amplification when system-reservoir interactions permit only single-quantum exchange. A controlled weak-nonlinearity expansion, supported by numerical
stationary solutions, shows that two-quantum exchange at the emitter enables amplification and predicts its onset. When two-quantum exchange dominates at the emitter and collector,
varying the base temperature redistributes populations between even and odd oscillator sectors. This changes the emitter and collector
heat currents far more than the base current, enabling large thermal amplification. We propose a realization using a longitudinally coupled auxiliary two-level system and bath spectral filtering, specifying conditions for its elimination to yield the required oscillator dynamics. These results establish nonlinear system-reservoir coupling
as a resource for controlling heat flow and enabling thermal amplification.
\end{abstract}

\maketitle

% =============================================================================
\section{Introduction}
\label{sec:introduction}
% =============================================================================

Controlling heat flow in quantum systems connects microscopic energy exchange with thermal management and thermal-device
operation~\cite{Goold2016,Vinjanampathy2016,Benenti2017,SegalAgarwalla2016,Naseem2024HeatCurrents}.
Electronic and superconducting circuits provide platforms for exploring this connection~\cite{Pekola2015,PekolaKarimi2021}.
Theoretical and experimental studies of phononic, electronic, and photonic heat conduction establish the role of individual transport
channels~\cite{Rego1998,Schwab2000,Jezouin2013,Meschke2006,Partanen2016}. Further control is demonstrated by phase-dependent heat transport and
photonic heat valves~\cite{Giazotto2012,Ronzani2018}, thermal rectifiers~\cite{MartinezPerez2015,Senior2020}, and three-terminal
superconducting circuits~\cite{Gubaydullin2022}. These developments motivate thermal components that regulate and process
heat currents~\cite{Li2012}. A thermal transistor provides a particularly useful function by allowing a third contact to control heat transport between two terminals.

Differential thermal amplification occurs when a base-temperature variation produces a transport-current change larger in magnitude
than the accompanying base-current change~\cite{Li2006,Joulain2016}. Related thermal-control functions span phononic logic, magnetic
transistors, electrically gated refrigeration, and radiative transistors~\cite{WangLi2007,Castelli2023,Saira2007,BenAbdallah2014}.
At the quantum level, interacting finite-level systems and quantum dots provide model platforms for transistor
operation~\cite{Mandarino2021,Rajapaksha2025,Zhang2018}. Superconducting and multimode electromechanical proposals connect this
function to device architectures~\cite{Majland2020,Nie2024}. Strong system-reservoir coupling can induce amplification and negative
differential thermal conductance~\cite{Wang2018,Liu2019}. However, negative differential thermal conductance is not a universal
requirement, as demonstrated theoretically in three-level and inelastic transport models~\cite{Wang2019,Jiang2015}.
Other approaches exploit optical control, periodic modulation, or detuning~\cite{Wijesekara2020,Gupt2022,Malavazi2024}.
Reservoir design offers further control through spectral filtering and engineered dissipation~\cite{Naseem2020HeatManager,Mandarino2022}, as
well as common, feedback-controlled, or collision-based environments~\cite{Liu2022,Ekanayake2024,Bhargava2026}.
These approaches motivate identifying how the exchange processes at the contacts contribute to amplification independently of the working
medium's intrinsic anharmonicity.

Reservoir engineering allows the exchange processes themselves to be designed, including channels that transfer quanta in
pairs~\cite{Poyatos1996,Verstraete2009,dxzd-nqrv}. Quadratic optomechanical coupling provides a route to two-phonon cooling
and squeezing~\cite{Nunnenkamp2010}, while nonlinear conversion through auxiliary lossy modes enables two-photon dissipation in superconducting
circuits~\cite{Leghtas2015,Marquet2024Autoparametric}. Two- and multiphoton driven dissipative processes have been proposed for
encoding and manipulating quantum information in oscillators~\cite{Mirrahimi2014}. Experiments using engineered two-photon dissipation demonstrate
suppression of cat-qubit bit flips, protected quantum control, and quantum error correction with concatenated bosonic
codes~\cite{Lescanne2020,Reglade2024,Putterman2025}. Applications also include quantum measurement and squeezed-state
preparation~\cite{Marquet2024}, together with theoretical proposals for photon blockade and thermal
rectification~\cite{Zhou2022,PhysRevA.105.012201, Ming2016, Naseem2026Rectification}. These applications motivate exploring engineered two-quantum exchange
as a resource for three-terminal heat control. Can such exchange at the reservoir contacts enable differential thermal
amplification while the working medium remains harmonic?

We address this question using a harmonic oscillator coupled to emitter, collector, and base reservoirs. Within a Lindblad master equation derived
under the Born--Markov and secular approximations, we recover the absence of amplification for single-quantum exchange alone, consistent with the
bound established for static harmonic networks~\cite{RieraCampeny2019}. Introducing two-quantum exchange at the emitter enables collector
amplification while the collector and base remain linearly coupled. A controlled weak-nonlinearity expansion predicts its onset, supported
by numerical stationary solutions. Allowing two-quantum exchange at both the emitter and collector produces
regimes of strong amplification. We explain this behavior by examining population dynamics without
single-quantum emitter and collector coupling. Two-quantum transitions preserve the even and odd excitation sectors,
while base-induced single-quantum transitions connect them. When two-quantum exchange dominates, the base temperature changes
the mean oscillator occupation mainly by redistributing populations between these sectors. At equal nonlinear rates, an exact identity
shows how this occupation change and the unequal reservoir occupations produce different emitter and collector current responses.
When their difference greatly exceeds the base-current response, large thermal amplification results. Finally, we propose a realization using a longitudinally coupled
auxiliary two-level system and bath spectral filtering, identifying the spectral and elimination conditions required to obtain the leading oscillator dynamics.

The remainder of the paper is organized as follows. Section~\ref{sec:model_formalism} introduces the model, reduced dynamics,
heat currents, and amplification measures. Section~\ref{sec:thermal_amplification} reviews the linear
no-amplification benchmark and analyzes two-quantum exchange at one and both transport contacts, explaining the resulting amplification
mechanism. Section~\ref{sec:experimental_realization} presents the proposed realization, and Sec.~\ref{sec:conclusions} summarizes the findings.

% =============================================================================
\section{Model and thermal-transistor formalism}
\label{sec:model_formalism}
% =============================================================================

We consider a three-terminal quantum thermal transistor consisting of
a single harmonic oscillator coupled to independent emitter (\(E\)),
collector (\(C\)), and base (\(B\)) reservoirs.
The emitter and collector serve as transport terminals, while the base
temperature \(T_B\) controls the stationary heat currents.
All three reservoirs can exchange single quanta with the oscillator,
and the transport contacts can additionally support two-quantum
exchange, while the base remains purely linear.
By retaining a harmonic oscillator Hamiltonian throughout, the model
isolates the role of nonlinear system-reservoir coupling in thermal
amplification without introducing intrinsic anharmonicity.
We first specify the microscopic interactions and derive the reduced
dynamics. We then define the bath-resolved heat currents and introduce the differential amplification
measures used to characterize transistor operation.

%-------------------------------------------------------------------------------
\subsection{Microscopic model and reduced master equation}
\label{subsec:microscopic_model_master_equation}
%-------------------------------------------------------------------------------

\begin{figure}[t]
    \centering
    \includegraphics[width=\columnwidth]{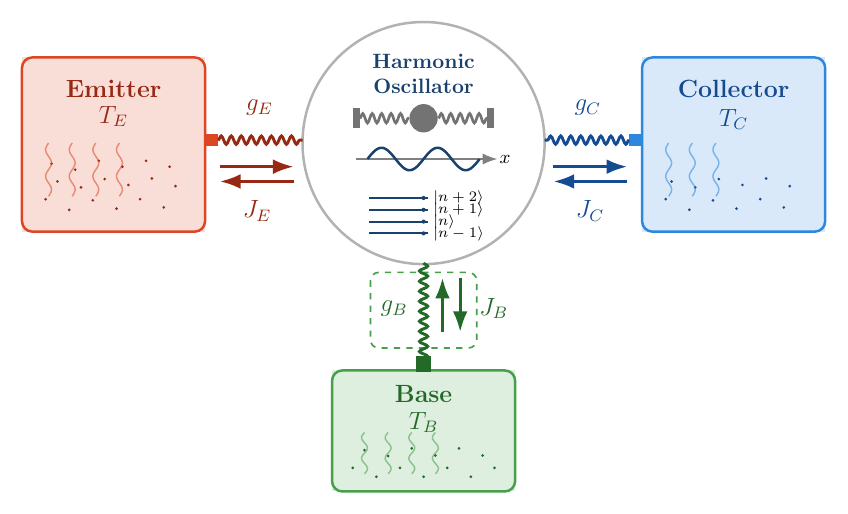}
    \caption{
    Schematic of the three-terminal thermal transistor. A harmonic
    oscillator is coupled to emitter (\(E\)), collector (\(C\)), and
    base (\(B\)) reservoirs at temperatures \(T_E\), \(T_C\), and \(T_B\),
    respectively. The labels \(J_E\), \(J_C\), and \(J_B\) identify the
    bath-resolved heat currents, with the stationary-state subscript
    omitted in the schematic. The central region shows the equally
    spaced oscillator Fock levels.
    }
    \label{fig:model_schematic}
\end{figure}

The three-terminal configuration shown in
Fig.~\ref{fig:model_schematic} is described by the total Hamiltonian
\begin{equation}
H
=
H_S+H_R+H_{SR},
\label{eq:total_hamiltonian}
\end{equation}
where \(H_S\), \(H_R\), and \(H_{SR}\) describe the oscillator, the
reservoirs, and their interaction, respectively. In units where
\(\hbar=k_B=1\), the oscillator Hamiltonian is
\begin{equation}
H_S
=
\omega a^\dagger a,
\label{eq:system_hamiltonian}
\end{equation}
where \(\omega\) is the oscillator frequency and \(a\) is its
annihilation operator. The reservoir Hamiltonian is
\begin{equation}
H_R
=
\sum_{\alpha=E,C,B}H_\alpha,
\qquad
H_\alpha
=
\sum_k
\omega_{\alpha k}
b_{\alpha k}^\dagger b_{\alpha k},
\label{eq:reservoir_hamiltonian}
\end{equation}
where \(\alpha\in\{E,C,B\}\) labels the emitter, collector, and base
reservoirs. The operator \(b_{\alpha k}\) annihilates an excitation of
frequency \(\omega_{\alpha k}\) in mode \(k\) of reservoir \(\alpha\).

To distinguish the two exchange channels, we decompose the
system-reservoir interaction as
\begin{equation}
H_{SR}
=
H_{SR}^{(1)}
+
H_{SR}^{(2)},
\label{eq:interaction_decomposition}
\end{equation}
where the superscripts \(1\) and \(2\) denote one-quantum and
two-quantum exchange, respectively. We adopt an energy-exchange
interaction at each contact. The linear contribution is
\begin{equation}
H_{SR}^{(1)}
=
\sum_{\alpha=E,C,B}
\left(
a^\dagger X_\alpha^{(1)}
+
aX_\alpha^{(1)\dagger}
\right),
\label{eq:linear_interaction}
\end{equation}
where \(X_\alpha^{(1)}\) is the reservoir coupling operator,
\begin{equation}
X_\alpha^{(1)}
=
\sum_k
g_{\alpha k}^{(1)}b_{\alpha k}.
\label{eq:linear_bath_coupling_operator}
\end{equation}
Here, \(g_{\alpha k}^{(1)}\) is the coupling amplitude between the
oscillator and mode \(k\) of reservoir \(\alpha\). This interaction
connects oscillator Fock states differing by one excitation and
exchanges an energy \(\omega\). The nonlinear contribution is restricted to the transport terminals
and takes the two-quantum exchange form~\cite{Voje2013}
\begin{equation}
H_{SR}^{(2)}
=
\sum_{\nu=E,C}
\left(
a^{\dagger 2}X_\nu^{(2)}
+
a^2X_\nu^{(2)\dagger}
\right),
\label{eq:nonlinear_interaction}
\end{equation}
where \(\nu\in\{E,C\}\) labels the emitter and collector, and
\(X_\nu^{(2)}\) is the corresponding reservoir coupling operator,
\begin{equation}
X_\nu^{(2)}
=
\sum_k
g_{\nu k}^{(2)}b_{\nu k}.
\label{eq:nonlinear_bath_coupling_operator}
\end{equation}
Here, \(g_{\nu k}^{(2)}\) is the coupling amplitude for two-quantum
oscillator exchange with mode \(k\) of reservoir \(\nu\). This
interaction connects oscillator Fock states differing by two
excitations and exchanges an energy \(2\omega\). The base contact has
no corresponding nonlinear contribution. A route for engineering the
effective nonlinear contacts is discussed in
Sec.~\ref{sec:experimental_realization}.

Each reservoir is described by a thermal state,
\begin{equation}
\rho_\alpha
=
\frac{
e^{-H_\alpha/T_\alpha}
}{
\operatorname{Tr}\!\left(e^{-H_\alpha/T_\alpha}\right)
},
\label{eq:thermal_reservoir_state}
\end{equation}
where \(T_\alpha\) is the temperature of reservoir \(\alpha\).
The temperatures are held constant during each dynamical evolution.
To characterize transistor operation, we vary \(T_B\) between
operating points while keeping \(T_E\), \(T_C\), and the prescribed
dissipative rates fixed. The reservoirs are mutually independent,
so their joint state is
\(\rho_R=\rho_E\otimes\rho_C\otimes\rho_B\),
where \(\rho_R\) denotes the combined reservoir state. The thermal
occupation entering the exchange rates is
\begin{equation}
\bar n_\alpha(\Omega)
=
\frac{1}{e^{\Omega/T_\alpha}-1},
\label{eq:bose_occupation}
\end{equation}
where \(\Omega>0\) is the frequency at which the reservoir is sampled.
For the one-quantum and two-quantum channels, respectively, we use
\begin{equation}
\bar n_\alpha
=
\bar n_\alpha(\omega),
\qquad
\bar m_\nu
=
\bar n_\nu(2\omega).
\label{eq:channel_thermal_occupations}
\end{equation}
Thus, \(\bar n_\alpha\) and \(\bar m_\nu\) are the reservoir
occupations associated with exchange energies \(\omega\) and
\(2\omega\).

To obtain the reduced dynamics, we assume an initially factorized state,
\(\rho_{\mathrm{tot}}(0)=\rho(0)\otimes\rho_R\), where
\(\rho_{\mathrm{tot}}\) is the joint oscillator-reservoir state and
\(\rho(t)\) is the reduced oscillator state.
Within the Born approximation, we retain terms to second order in
the microscopic coupling amplitudes and treat the reservoirs as
remaining in their stationary thermal states.
The Markov approximation requires reservoir correlation times to be
short compared with the relevant oscillator relaxation times.
The interactions in Eqs.~\eqref{eq:linear_interaction} and
\eqref{eq:nonlinear_interaction} connect transitions at \(\omega\)
and \(2\omega\).
We apply the secular approximation between these frequency sectors,
requiring their separation to be large compared with the relevant
dissipative linewidths.
We neglect reservoir-induced Lamb shifts and retain the harmonic
Hamiltonian in Eq.~\eqref{eq:system_hamiltonian}.
Tracing over the reservoirs under these assumptions gives~\cite{BreuerPetruccione2002}
\begin{equation}
\frac{d\rho(t)}{dt}
=
-i[H_S,\rho(t)]
+
\sum_{\alpha=E,C,B}
\mathcal L_\alpha^{(1)}\rho(t)
+
\sum_{\nu=E,C}
\mathcal L_\nu^{(2)}\rho(t).
\label{eq:general_master_equation}
\end{equation}
Here, \(\mathcal L_\alpha^{(1)}\) and
\(\mathcal L_\nu^{(2)}\) are the one-quantum and two-quantum
dissipative generators, respectively.
Evaluating the reservoir correlations at \(\omega\) gives
\begin{equation}
\mathcal L_\alpha^{(1)}\rho(t)
=
\gamma_\alpha(\bar n_\alpha+1)
\mathcal D[a]\rho(t)
+
\gamma_\alpha\bar n_\alpha
\mathcal D[a^\dagger]\rho(t),
\label{eq:linear_dissipator}
\end{equation}
where \(\gamma_\alpha\) is the one-quantum coupling rate, with
\(\gamma_E,\gamma_C\geq0\) and \(\gamma_B>0\).
The Lindblad dissipator is defined by
\begin{equation}
\mathcal D[O]\rho(t)
=
O\rho(t)O^\dagger
-
\frac{1}{2}
\left\{
O^\dagger O,\rho(t)
\right\},
\label{eq:lindblad_dissipator}
\end{equation}
where \(O\) is the jump operator and
\(\{A,B\}=AB+BA\) denotes the anticommutator.
The two terms in Eq.~\eqref{eq:linear_dissipator} describe
single-quantum relaxation and excitation, respectively.
Similarly, the reservoir correlations at \(2\omega\) yield
\begin{equation}
\mathcal L_\nu^{(2)}\rho(t)
=
\Gamma_\nu(\bar m_\nu+1)
\mathcal D[a^2]\rho(t)
+
\Gamma_\nu\bar m_\nu
\mathcal D[a^{\dagger 2}]\rho(t),
\label{eq:nonlinear_dissipator}
\end{equation}
where \(\Gamma_\nu\geq0\) is the two-quantum coupling rate. The two terms describe two-quantum relaxation and excitation,
respectively. With the reduced dynamics established, we next define the bath-resolved heat currents and the differential
responses used to characterize thermal amplification.

%-------------------------------------------------------------------------------
\subsection{Heat currents and thermal amplification}
\label{subsec:heat_currents_amplification}
%-------------------------------------------------------------------------------

Throughout this work, we analyze the thermal transistor in the steady
state. To distinguish the dynamical energy balance from its stationary
limit, we first define the time-dependent heat current associated with
reservoir \(\alpha\)~\cite{Kosloff2013},
\begin{equation}
J_\alpha(t)
=
\operatorname{Tr}
\!\left[
H_S\mathcal L_\alpha\rho(t)
\right].
\label{eq:time_heat_current_definition}
\end{equation}
Here, \(J_\alpha(t)>0\) denotes energy flow from reservoir \(\alpha\)
into the oscillator, and \(\mathcal L_\alpha\) is the complete
dissipative generator associated with that reservoir. For the transport
terminals \(\nu\in\{E,C\}\), it contains both one-quantum and two-quantum
contributions,
\(\mathcal L_\nu=\mathcal L_\nu^{(1)}+\mathcal L_\nu^{(2)}\),
whereas the base contains only the one-quantum contribution,
\(\mathcal L_B=\mathcal L_B^{(1)}\).

The normalized stationary state \(\rho_{\mathrm{ss}}\) is obtained
by setting the time derivative in
Eq.~\eqref{eq:general_master_equation} to zero. Evaluating the
heat current in this state gives
\begin{equation}
J_{\alpha,\mathrm{ss}}
=
\operatorname{Tr}
\!\left[
H_S\mathcal L_\alpha\rho_{\mathrm{ss}}
\right],
\label{eq:heat_current_definition}
\end{equation}
where \(J_{\alpha,\mathrm{ss}}>0\) retains the same convention of
energy flow into the oscillator. Multiplying Eq.~\eqref{eq:general_master_equation} by \(H_S\),
taking the trace, and using the current definition in
Eq.~\eqref{eq:time_heat_current_definition} gives
\[
\frac{d}{dt}
\operatorname{Tr}\!\left[H_S\rho(t)\right]
=
J_E(t)+J_C(t)+J_B(t).
\]
Here, \(H_S\) is time independent, and the Hamiltonian contribution
vanishes by cyclicity of the trace,
\(\operatorname{Tr}\!\left(H_S[H_S,\rho(t)]\right)=0\).
At steady state, the left-hand side vanishes, giving
\begin{equation}
J_{E,\mathrm{ss}} + J_{C,\mathrm{ss}} + J_{B,\mathrm{ss}} = 0.
\label{eq:steady_state_energy_conservation}
\end{equation}
Equation~\eqref{eq:steady_state_energy_conservation} expresses
steady-state energy conservation, requiring the net heat current
entering the oscillator to vanish.

Following the control protocol specified in
Sec.~\ref{subsec:microscopic_model_master_equation}, we vary \(T_B\)
while keeping \(T_E\), \(T_C\), \(\omega\), and all dissipative rates
fixed. The response between neighboring stationary operating points
is quantified by
\begin{equation}
\kappa_\alpha
=
\frac{\partial J_{\alpha,\mathrm{ss}}}{\partial T_B},
\label{eq:differential_thermal_response}
\end{equation}
where \(\kappa_\alpha\) is the differential thermal response of the
stationary current associated with reservoir \(\alpha\). Differentiating
Eq.~\eqref{eq:steady_state_energy_conservation} with respect to \(T_B\)
gives
\begin{equation}
\kappa_E+\kappa_C+\kappa_B=0.
\label{eq:differential_current_conservation}
\end{equation}
To compare the transport response with the base response while
retaining their relative signs, we define
\begin{equation}
\beta_\nu
=
\frac{
\partial_{T_B}J_{\nu,\mathrm{ss}}
}{
\partial_{T_B}J_{B,\mathrm{ss}}
}
=
\frac{\kappa_\nu}{\kappa_B},
\qquad
\kappa_B\neq0,
\label{eq:signed_thermal_gain}
\end{equation}
where \(\beta_\nu\) is the signed differential gain of transport
terminal \(\nu\). The second equality follows from
Eq.~\eqref{eq:differential_thermal_response}. The magnitude of the signed
gain defines the thermal amplification factor,
\begin{equation}
\mathcal A_\nu
=
|\beta_\nu|,
\label{eq:thermal_amplification_factor}
\end{equation}
where \(\mathcal A_\nu\) compares the magnitudes of the transport and
base current responses. Dividing
Eq.~\eqref{eq:differential_current_conservation} by \(\kappa_B\) and
using Eq.~\eqref{eq:signed_thermal_gain} yields
\begin{equation}
\beta_E+\beta_C=-1.
\label{eq:signed_gain_conservation}
\end{equation}
Thermal amplification occurs when
\begin{equation}
\mathcal A_E>1
\qquad\text{or}\qquad
\mathcal A_C>1.
\label{eq:thermal_amplification_condition}
\end{equation}
This condition means that a small variation in \(T_B\) produces a
larger current change in magnitude at one or both transport terminals
than at the base.

% =============================================================================
\section{Thermal amplification}
\label{sec:thermal_amplification}
% =============================================================================

We now examine whether nonlinear environmental coupling enables thermal
amplification in the harmonic working medium. We first derive the
stationary properties of the fully linear benchmark and recover its
absence of amplification. We then introduce two-quantum exchange at
the emitter and use a controlled weak-nonlinearity expansion to
determine the onset of amplification. Finally, we include two-quantum
exchange at both the emitter and collector and analyze the
differential responses underlying strong amplification.

%-------------------------------------------------------------------------------
\subsection{Fully linear coupling and absence of thermal amplification}
\label{subsec:linear_benchmark}
%-------------------------------------------------------------------------------

We first consider the fully linear benchmark, recovering the absence
of differential thermal amplification established for static harmonic
networks~\cite{RieraCampeny2019}. We derive the single-oscillator
expressions needed for the subsequent nonlinear analysis.
This configuration follows from
Eq.~\eqref{eq:general_master_equation} by setting
\(\Gamma_E=\Gamma_C=0\). To determine its population dynamics,
we substitute the dissipators in Eq.~\eqref{eq:linear_dissipator}
into the master equation~\eqref{eq:general_master_equation} and take diagonal matrix elements in the
oscillator Fock basis. Using the oscillator ladder-operator matrix
elements gives
\begin{align}
\frac{dp_j^{(0)}(t)}{dt}
={}&
(S+\gamma_\Sigma)
\left[
(j+1)p_{j+1}^{(0)}(t)
-
jp_j^{(0)}(t)
\right]
\nonumber\\
&+
S
\left[
jp_{j-1}^{(0)}(t)
-
(j+1)p_j^{(0)}(t)
\right].
\label{eq:linear_population_equation}
\end{align}
Here,
\(p_j^{(0)}(t)=\langle j|\rho^{(0)}(t)|j\rangle\)
is the time-dependent population of Fock level \(j=0,1,\ldots\),
and the superscript \((0)\) denotes the fully linear reference
configuration. The coefficients appearing in
Eq.~\eqref{eq:linear_population_equation} are
\begin{align}
\gamma_\Sigma
&=
\gamma_E+\gamma_C+\gamma_B,
\label{eq:linear_total_rate}
\\
S
&=
\gamma_E\bar n_E
+
\gamma_C\bar n_C
+
\gamma_B\bar n_B,
\label{eq:linear_weighted_occupation_rate}
\end{align}
where \(\gamma_\Sigma\) is the total linear relaxation rate and
\(S\) is the thermally induced excitation coefficient.
The populations satisfy \(p_{-1}^{(0)}(t)=0\) and
\(\sum_{j=0}^{\infty}p_j^{(0)}(t)=1\).

Equation~\eqref{eq:linear_population_equation} describes exchange
between adjacent oscillator levels. At steady state, we set its
left-hand side to zero and write
\(p_{j,\mathrm{ss}}^{(0)}
=\langle j|\rho_{\mathrm{ss}}^{(0)}|j\rangle\)
for the stationary populations. The net probability current between
neighboring levels is then constant. The ground-state boundary
requires this current to vanish, giving
\begin{equation}
p_{j+1,\mathrm{ss}}^{(0)}
=
r\,p_{j,\mathrm{ss}}^{(0)},
\qquad
r
=
\frac{S}{S+\gamma_\Sigma},
\label{eq:linear_stationary_recurrence}
\end{equation}
where \(r\) is the ratio of successive stationary populations when
those populations are nonzero. For finite reservoir occupations and
positive \(\gamma_\Sigma\), \(0\leq r<1\). Iterating
Eq.~\eqref{eq:linear_stationary_recurrence} and imposing normalization
gives
\begin{equation}
p_{j,\mathrm{ss}}^{(0)}
=
(1-r)r^j.
\label{eq:linear_stationary_distribution}
\end{equation}
This geometric distribution is the exact stationary population
distribution of the fully linear model.
The linear heat currents depend on the stationary mean occupation.
Evaluating the first moment of
Eq.~\eqref{eq:linear_stationary_distribution} gives
\begin{equation}
n_{\mathrm{ss}}^{(0)}
=
\sum_{j=0}^{\infty}j\,p_{j,\mathrm{ss}}^{(0)}
=
\frac{S}{\gamma_\Sigma},
\label{eq:linear_stationary_occupation}
\end{equation}
where \(n_{\mathrm{ss}}^{(0)}\) is the mean oscillator occupation
in the fully linear stationary state. It is the coupling-weighted
average of the three reservoir occupations. Substituting the linear
dissipator in Eq.~\eqref{eq:linear_dissipator} into the heat-current
definition in Eq.~\eqref{eq:heat_current_definition}, using the
oscillator Hamiltonian \(H_S\), and evaluating the trace gives
\begin{equation}
J_{\alpha,\mathrm{ss}}^{(0)}
=
\omega\gamma_\alpha
\left(
\bar n_\alpha-n_{\mathrm{ss}}^{(0)}
\right).
\label{eq:linear_heat_currents}
\end{equation}
The current is positive when the reservoir occupation exceeds the
stationary oscillator occupation. Summing
Eq.~\eqref{eq:linear_heat_currents} over the reservoirs gives
\(\sum_\alpha J_{\alpha,\mathrm{ss}}^{(0)}
=\omega(S-\gamma_\Sigma n_{\mathrm{ss}}^{(0)})=0\).
The currents therefore satisfy energy conservation in
Eq.~\eqref{eq:steady_state_energy_conservation}.

To assess thermal amplification, we differentiate the stationary
currents in Eq.~\eqref{eq:linear_heat_currents} with respect to \(T_B\).
Using Eq.~\eqref{eq:linear_stationary_occupation}, with only the base
occupation varying, gives
\begin{align}
\kappa_E^{(0)}
&=
-\omega
\frac{\gamma_E\gamma_B}{\gamma_\Sigma}
\bar n_B',
\label{eq:linear_emitter_response}
\\
\kappa_C^{(0)}
&=
-\omega
\frac{\gamma_C\gamma_B}{\gamma_\Sigma}
\bar n_B',
\label{eq:linear_collector_response}
\\
\kappa_B^{(0)}
&=
\omega
\frac{\gamma_B(\gamma_E+\gamma_C)}{\gamma_\Sigma}
\bar n_B'.
\label{eq:linear_base_response}
\end{align}
Here, the prime denotes differentiation with respect to \(T_B\).
Differentiating the Bose occupation in Eq.~\eqref{eq:bose_occupation}
at \(\Omega=\omega\) yields
\begin{equation}
\bar n_B'
=
\frac{\partial\bar n_B}{\partial T_B}
=
\frac{\omega}{T_B^2}
\bar n_B(\bar n_B+1).
\label{eq:base_occupation_derivative}
\end{equation}
For finite \(T_B>0\) and positive coupling rates,
\(\kappa_B^{(0)}>0\), so the differential gains are well defined.
Substituting Eqs.~\eqref{eq:linear_emitter_response}--%
\eqref{eq:linear_base_response} into
Eq.~\eqref{eq:signed_thermal_gain} gives
\begin{equation}
\beta_E^{(0)}
=
-\frac{\gamma_E}{\gamma_E+\gamma_C},
\qquad
\beta_C^{(0)}
=
-\frac{\gamma_C}{\gamma_E+\gamma_C}.
\label{eq:linear_signed_gains}
\end{equation}
Taking their magnitudes gives the amplification factors,
\begin{equation}
\mathcal A_E^{(0)}
=
\frac{\gamma_E}{\gamma_E+\gamma_C},
\qquad
\mathcal A_C^{(0)}
=
\frac{\gamma_C}{\gamma_E+\gamma_C}.
\label{eq:linear_amplification_factors}
\end{equation}
For positive emitter and collector couplings,
\begin{equation}
0<\mathcal A_E^{(0)}<1,
\qquad
0<\mathcal A_C^{(0)}<1,
\qquad
\mathcal A_E^{(0)}+\mathcal A_C^{(0)}=1.
\label{eq:linear_no_amplification}
\end{equation}
These expressions recover the absence of amplification in the fully linear configuration: neither terminal satisfies the condition in
Eq.~\eqref{eq:thermal_amplification_condition}.

The linear gains in Eq.~\eqref{eq:linear_signed_gains} depend only on the relative emitter and collector coupling rates.
The reservoir temperatures and base coupling affect the currents and their responses but cancel from the gain ratios.
Physically, varying \(T_B\) changes the oscillator occupation, giving both transport currents responses of the same
sign, weighted by their respective coupling rates. Their magnitudes sum to the base response, so neither can exceed it.
We next introduce two-quantum exchange at the emitter while keeping the linear parameters unchanged to examine how nonlinear
system-reservoir coupling modifies this response and enables thermal amplification.

% -----------------------------------------------------------------------------
\subsection{Nonlinear coupling at the emitter}
\label{subsec:nonlinear_emitter}
% -----------------------------------------------------------------------------

The exact linear result in Eq.~\eqref{eq:linear_no_amplification} shows
that purely linear reservoir coupling cannot produce differential thermal
amplification. We therefore activate the two-quantum channel only at the
emitter, with \(\Gamma_E>0\) and \(\Gamma_C=0\), while keeping the
oscillator Hamiltonian and all linear parameters unchanged. We first
derive the population and moment equations and identify their stationary
forms, which provide the starting point for a controlled weak-nonlinearity
expansion to determine the onset of amplification.

% .............................................................................
\subsubsection{Exact emitter-only equations and moment structure}
\label{subsubsec:nonlinear_emitter_exact_equations}
% .............................................................................

Substituting the linear dissipators and the nonlinear emitter dissipator
from Eqs.~\eqref{eq:linear_dissipator} and
\eqref{eq:nonlinear_dissipator} into
Eq.~\eqref{eq:general_master_equation} and taking diagonal Fock-basis
matrix elements gives
\begin{equation}
\begin{aligned}
\frac{dp_j(t)}{dt}
={}&(S+\gamma_\Sigma)
\bigl[(j+1)p_{j+1}-jp_j\bigr]
\\
&+S\bigl[jp_{j-1}-(j+1)p_j\bigr]
\\
&+\Gamma_E(\bar m_E+1)
\bigl[(j+2)(j+1)p_{j+2}-j(j-1)p_j\bigr]
\\
&+\Gamma_E\bar m_E
\bigl[j(j-1)p_{j-2}-(j+1)(j+2)p_j\bigr].
\end{aligned}
\label{eq:emitter_nonlinear_population_equation}
\end{equation}
Here, \(p_j(t)=\langle j|\rho(t)|j\rangle\) is the Fock-state
population, with time arguments suppressed on the right-hand side. 
In Eq.~\eqref{eq:emitter_nonlinear_population_equation}, the first two lines describe single-quantum relaxation and excitation,
reproducing the linear population dynamics in Eq.~\eqref{eq:linear_population_equation}.
The last two lines describe two-quantum relaxation and excitation at the emitter, connecting Fock states differing by two excitations.
At steady state, we set the left-hand side of Eq.~\eqref{eq:emitter_nonlinear_population_equation} to zero and write
\(p_{j,\mathrm{ss}}=\langle j|\rho_{\mathrm{ss}}|j\rangle\) for the stationary populations.

We next obtain the corresponding stationary heat currents. Substituting
the linear dissipator in Eq.~\eqref{eq:linear_dissipator} into the
heat-current definition in Eq.~\eqref{eq:heat_current_definition}
and evaluating the trace gives
\begin{equation}
\begin{aligned}
J_{C,\mathrm{ss}}
&=
\omega\gamma_C
\left(
\bar n_C-n_{\mathrm{ss}}
\right),
\\
J_{B,\mathrm{ss}}
&=
\omega\gamma_B
\left(
\bar n_B-n_{\mathrm{ss}}
\right),
\end{aligned}
\label{eq:emitter_nonlinear_collector_base_currents}
\end{equation}
where \(n_{\mathrm{ss}}\) is the stationary value of the
time-dependent mean oscillator occupation
\(n(t)=\operatorname{Tr}[a^\dagger a\rho(t)]\).
These currents retain the linear form in
Eq.~\eqref{eq:linear_heat_currents}, although the occupation is now
determined by the nonlinear stationary equations.
The emitter current contains both one-quantum and two-quantum
contributions. Substituting the dissipators in
Eqs.~\eqref{eq:linear_dissipator} and
\eqref{eq:nonlinear_dissipator} into
Eq.~\eqref{eq:heat_current_definition} and evaluating the trace gives
\begin{equation}
\begin{aligned}
J_{E,\mathrm{ss}}
={}&
\omega\gamma_E
\left(
\bar n_E-n_{\mathrm{ss}}
\right)
\\
&+
2\omega\Gamma_E
\left[
-M_{2,\mathrm{ss}}
+
4\bar m_E n_{\mathrm{ss}}
+
2\bar m_E
\right],
\end{aligned}
\label{eq:emitter_nonlinear_emitter_current}
\end{equation}
where \(M_{2,\mathrm{ss}}\) is the stationary value of the second
factorial moment
\(M_2(t)=\operatorname{Tr}[a^{\dagger 2}a^2\rho(t)]\).
In Eq. \eqref{eq:emitter_nonlinear_emitter_current}, the factor \(2\omega\) accounts for the energy exchanged in each
two-quantum transition. Unlike the collector and base currents, the emitter current depends explicitly on both \(n_{\mathrm{ss}}\) and
\(M_{2,\mathrm{ss}}\).

To evaluate the stationary currents analytically, we must determine
these two moments. Multiplying
Eq.~\eqref{eq:emitter_nonlinear_population_equation} by \(j\),
summing over the Fock levels, and shifting the summation indices gives
\begin{equation}
\begin{aligned}
\frac{dn(t)}{dt}
={}&
S-\gamma_\Sigma n(t)
\\
&+
2\Gamma_E
\left[
-M_2(t)
+
4\bar m_E n(t)
+
2\bar m_E
\right].
\end{aligned}
\label{eq:emitter_nonlinear_number_equation}
\end{equation}
At steady state, the left-hand side is zero, and \(n(t)\) and
\(M_2(t)\) take the values \(n_{\mathrm{ss}}\) and
\(M_{2,\mathrm{ss}}\), respectively.
Unlike the fully linear case, the stationary form of
Eq.~\eqref{eq:emitter_nonlinear_number_equation} does not determine
\(n_{\mathrm{ss}}\) alone because it also contains
\(M_{2,\mathrm{ss}}\).
To determine whether these moments form a closed set, we multiply
Eq.~\eqref{eq:emitter_nonlinear_population_equation} by \(j(j-1)\)
and sum over the Fock levels. This gives
\begin{align}
\frac{dM_2(t)}{dt}
={}&
-2\gamma_\Sigma M_2(t)
+
4Sn(t)
\nonumber\\
&+
\Gamma_E
\left[
-4M_3(t)
+
(24\bar m_E-2)M_2(t)
\right.
\nonumber\\
&\hspace{3.8em}\left.
+
32\bar m_E n(t)
+
4\bar m_E
\right],
\label{eq:emitter_nonlinear_second_moment_equation}
\end{align}
where \(M_3(t)=\sum_{j=0}^{\infty}j(j-1)(j-2)p_j(t)\) is the third factorial moment, with stationary value \(M_{3,\mathrm{ss}}\).
Setting the left-hand side of Eq.~\eqref{eq:emitter_nonlinear_second_moment_equation} to zero gives the corresponding stationary relation.
This relation introduces \(M_{3,\mathrm{ss}}\), whose stationary equation involves still higher factorial moments. The exact nonlinear
moment hierarchy therefore does not close at finite order.

A moment-closure approximation is unnecessary for the numerical calculation, since the stationary populations can be obtained directly
from the stationary form of Eq.~\eqref{eq:emitter_nonlinear_population_equation} using a
converged Fock-space truncation. Analytical expressions for the currents, however, require a controlled treatment of the higher moments.
The weak-nonlinearity regime permits an expansion around the known linear stationary state, which we develop next.

% .............................................................................
\subsubsection{Controlled weak-nonlinearity expansion and amplification threshold}
\label{subsubsec:nonlinear_emitter_weak}
% .............................................................................

We expand the moments in Eqs.~\eqref{eq:emitter_nonlinear_number_equation} and~\eqref{eq:emitter_nonlinear_second_moment_equation}  around the fully linear state in the parameter
\begin{equation}
\epsilon
=
\frac{\Gamma_E}{\gamma_\Sigma},
\qquad
\epsilon\ll1,
\label{eq:weak_nonlinearity_parameter}
\end{equation}
where \(\epsilon\) measures the nonlinear emitter rate relative to
the total linear relaxation rate.
At fixed reservoir temperatures and linear coupling rates, we write
\begin{align}
n_{\mathrm{ss}}
&=
n_{\mathrm{ss}}^{(0)}
+
\epsilon n_{\mathrm{ss}}^{(1)}
+
O(\epsilon^2),
\nonumber\\
M_{2,\mathrm{ss}}
&=
M_{2,\mathrm{ss}}^{(0)}
+
O(\epsilon),
\label{eq:weak_stationary_moment_expansion}
\end{align}
where \(n_{\mathrm{ss}}^{(1)}\) is the first-order occupation
coefficient and \(n_{\mathrm{ss}}^{(0)}\) is given in
Eq.~\eqref{eq:linear_stationary_occupation}.
The geometric distribution in
Eq.~\eqref{eq:linear_stationary_distribution} gives
\begin{equation}
M_{2,\mathrm{ss}}^{(0)}
=
\sum_{j=0}^{\infty}
j(j-1)p_{j,\mathrm{ss}}^{(0)}
=
2\left(n_{\mathrm{ss}}^{(0)}\right)^2.
\label{eq:weak_second_moment_zeroth_order}
\end{equation}
This is an exact property of the linear reference state.
In the stationary form of
Eq.~\eqref{eq:emitter_nonlinear_number_equation},
\(M_{2,\mathrm{ss}}\) is multiplied by
\(\Gamma_E=\epsilon\gamma_\Sigma\), so its first-order correction
contributes only at second order in \(\epsilon\).
Its zeroth-order value therefore suffices to determine the leading
occupation correction without imposing a moment closure.

Substituting Eqs.~\eqref{eq:weak_stationary_moment_expansion} and \eqref{eq:weak_second_moment_zeroth_order}
into the stationary form of Eq.~\eqref{eq:emitter_nonlinear_number_equation}, and using
\(S=\gamma_\Sigma n_{\mathrm{ss}}^{(0)}\), gives
\begin{equation}
n_{\mathrm{ss}}
=
n_{\mathrm{ss}}^{(0)}
+
4\epsilon Q_0
+
O(\epsilon^2),
\label{eq:weak_stationary_occupation}
\end{equation}
where
\begin{equation}
Q_0
=
\bar m_E
\left(2n_{\mathrm{ss}}^{(0)}+1\right)
-
\left(n_{\mathrm{ss}}^{(0)}\right)^2.
\label{eq:weak_Q0_definition}
\end{equation}
The quantity \(Q_0\) describes the net two-quantum excitation tendency
in the linear reference state.
Its sign determines whether nonlinear emitter exchange increases or
decreases the stationary occupation to first order.
Substituting Eqs.~\eqref{eq:weak_second_moment_zeroth_order}
and \eqref{eq:weak_stationary_occupation} into
Eqs.~\eqref{eq:emitter_nonlinear_collector_base_currents}
and \eqref{eq:emitter_nonlinear_emitter_current} yields
\begin{align}
J_{E,\mathrm{ss}}
&=
J_{E,\mathrm{ss}}^{(0)}
+
\frac{
4\omega\Gamma_E(\gamma_C+\gamma_B)
}{
\gamma_\Sigma
}
Q_0
+
O(\omega\gamma_\Sigma\epsilon^2),
\nonumber\\
J_{C,\mathrm{ss}}
&=
J_{C,\mathrm{ss}}^{(0)}
-
\frac{
4\omega\gamma_C\Gamma_E
}{
\gamma_\Sigma
}
Q_0
+
O(\omega\gamma_\Sigma\epsilon^2),
\nonumber\\
J_{B,\mathrm{ss}}
&=
J_{B,\mathrm{ss}}^{(0)}
-
\frac{
4\omega\gamma_B\Gamma_E
}{
\gamma_\Sigma
}
Q_0
+
O(\omega\gamma_\Sigma\epsilon^2).
\label{eq:weak_heat_currents}
\end{align}
Here, \(J_{\alpha,\mathrm{ss}}^{(0)}\) are the linear currents in
Eq.~\eqref{eq:linear_heat_currents}.
Thus, \(Q_0\) governs the leading corrections to all three currents.

To determine the differential current responses, we differentiate
Eq.~\eqref{eq:weak_Q0_definition} with respect to \(T_B\).
Holding \(\bar m_E\) fixed and using
Eqs.~\eqref{eq:linear_stationary_occupation} and
\eqref{eq:base_occupation_derivative} gives
\begin{equation}
\frac{\partial Q_0}{\partial T_B}
=
2\Delta
\frac{\gamma_B}{\gamma_\Sigma}
\bar n_B',
\label{eq:weak_Q0_temperature_response}
\end{equation}
where
\begin{equation}
\Delta
=
\bar m_E-n_{\mathrm{ss}}^{(0)}.
\label{eq:weak_occupation_difference}
\end{equation}
While \(Q_0\) determines the current corrections, \(\Delta\) controls their response to the base temperature.
Differentiating Eq.~\eqref{eq:weak_heat_currents} and expanding the gain ratios in Eq.~\eqref{eq:signed_thermal_gain} consistently
to first order gives
\begin{align}
\beta_C
&=
-\frac{\gamma_C}{\gamma_E+\gamma_C}
-
\frac{
8\gamma_C\Gamma_E
}{
(\gamma_E+\gamma_C)^2
}
\Delta
+
O(\epsilon^2),
\nonumber\\
\beta_E
&=
-\frac{\gamma_E}{\gamma_E+\gamma_C}
+
\frac{
8\gamma_C\Gamma_E
}{
(\gamma_E+\gamma_C)^2
}
\Delta
+
O(\epsilon^2).
\label{eq:weak_signed_gains}
\end{align}
For \(\Delta>0\), the collector gain becomes more negative,
increasing its magnitude toward the amplification threshold.
The gain expansion also requires control of the correction to its
denominator.
Differentiating the base current in
Eq.~\eqref{eq:weak_heat_currents} and normalizing by
Eq.~\eqref{eq:linear_base_response} gives
\begin{equation}
\frac{\kappa_B}{\kappa_B^{(0)}}
=
1
-
\frac{8\epsilon\gamma_B}{\gamma_E+\gamma_C}\Delta
+
O(\epsilon^2).
\label{eq:weak_base_response}
\end{equation}
The first-order correction in this ratio must remain small.
Thus, \(\epsilon\ll1\) alone does not ensure a reliable gain
expansion for arbitrary temperatures and coupling ratios.
The neglected higher-order corrections to the stationary moments
must also remain negligible.

On the branch \(\beta_C<0\), collector amplification requires
\(\beta_C<-1\).
Setting the first-order collector gain in
Eq.~\eqref{eq:weak_signed_gains} equal to \(-1\) gives
\begin{equation}
\Gamma_{E,\mathrm{th}}^{(1)}(T_B)
=
\frac{
\gamma_E(\gamma_E+\gamma_C)
}{
8\gamma_C
\left[
\bar m_E-n_{\mathrm{ss}}^{(0)}(T_B)
\right]
},
\qquad
\Delta(T_B)>0,
\label{eq:weak_amplification_threshold}
\end{equation}
where \(\Gamma_{E,\mathrm{th}}^{(1)}\) is the first-order estimate
of the emitter two-quantum rate at the onset of amplification.
Its temperature dependence enters through
\(n_{\mathrm{ss}}^{(0)}(T_B)\).
Collector amplification is predicted for
\(\Gamma_E>\Gamma_{E,\mathrm{th}}^{(1)}(T_B)\), provided the
expansion remains valid.
% -----------------------------------------------------------------------------
\begin{figure}[t]
    \centering
    \includegraphics[width=\columnwidth]{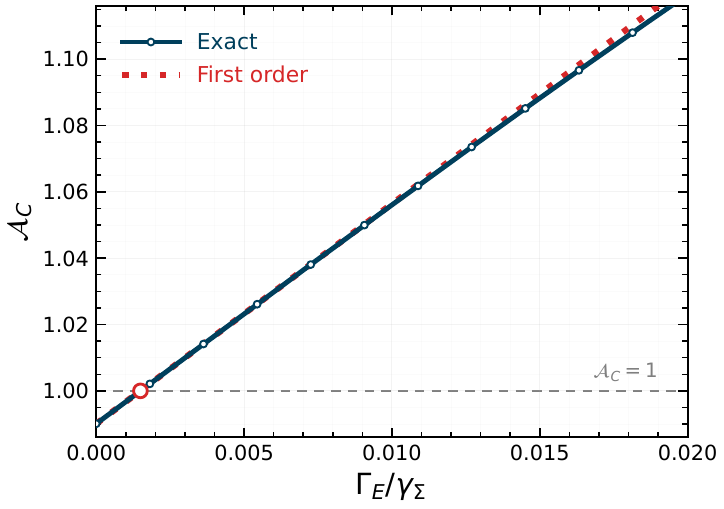}
    \caption{
    Collector thermal amplification \(\mathcal A_C\) as a function
    of the scaled emitter two-quantum rate
    \(\Gamma_E/\gamma_\Sigma\), with \(\Gamma_C=0\).
    The solid blue curve with markers is obtained by solving the
    stationary form of
    Eq.~\eqref{eq:emitter_nonlinear_population_equation}.
    The red dotted curve is the magnitude of the first-order collector
    gain in Eq.~\eqref{eq:weak_signed_gains}.
    The horizontal dashed line marks the onset criterion
    \(\mathcal A_C=1\).
    The parameters are
    \(T_E/\omega=2.0\), \(T_C/\omega=0.1\),
    \(T_B/\omega=0.3\),
    \(\gamma_E/\omega=1.0\times10^{-3}\),
    \(\gamma_C/\omega=1.0\times10^{-1}\), and
    \(\gamma_B/\omega=5.0\times10^{-2}\), with
    \(\hbar=k_B=1\).
    }
    \label{fig:weak_nonlinearity_crossover}
\end{figure}
% -----------------------------------------------------------------------------
Figure~\ref{fig:weak_nonlinearity_crossover} compares the first-order collector amplification \(\mathcal A_C\), obtained by
taking the magnitude of the collector gain in Eq.~\eqref{eq:weak_signed_gains}, with the exact numerical result
from the stationary form of Eq.~\eqref{eq:emitter_nonlinear_population_equation}.
The analytical and numerical curves agree closely over the displayed interval, supporting the first-order prediction
for the amplification magnitude. The threshold predicted by Eq.~\eqref{eq:weak_amplification_threshold} also agrees closely
with the nonlinear rate at which the numerical curve crosses \(\mathcal A_C=1\).

The competing roles of the emitter channels are evident from Eq.~\eqref{eq:weak_signed_gains}. For \(\Delta>0\), increasing
\(\Gamma_E\) makes the collector gain more negative and therefore increases its magnitude within the weak-nonlinearity regime.
Setting \(\gamma_E=0\) removes the negative linear contribution to \(\beta_E\), leaving a positive first-order correction.
Since \(\beta_E+\beta_C=-1\), a positive \(\beta_E\) implies \(\beta_C<-1\), and hence collector amplification.
Thus, arbitrarily weak positive two-quantum coupling can enable amplification.
For finite \(\gamma_E\), the nonlinear correction must first overcome the negative linear contribution.
This identifies nonlinear system-reservoir coupling through two-quantum exchange as the mechanism enabling amplification, while the single-quantum
channel opposes it.

% -----------------------------------------------------------------------------
\subsection{Nonlinear coupling at the emitter and collector}
\label{subsec:nonlinear_emitter_collector}
% -----------------------------------------------------------------------------

Section~\ref{subsec:nonlinear_emitter} establishes that two-quantum
exchange at the emitter can enable thermal amplification. We now
allow two-quantum exchange at both the emitter and collector,
with independent nonlinear coupling rates, while the base remains
linearly coupled. We examine stationary transport beyond the
weak-nonlinearity expansion by varying these two rates across
configurations. At each configuration, the differential responses
are evaluated by varying \(T_B\) with all coupling rates,
\(T_E\), \(T_C\), and \(\omega\) held fixed. We first establish
amplification numerically, then consider equal nonlinear rates
to explain its mechanism and examine how single-quantum
transport exchange modifies it.

% .............................................................................
\subsubsection{Stationary transport and thermal amplification}
\label{subsubsec:nonlinear_emitter_collector_exact}
% .............................................................................

% -----------------------------------------------------------------------------
\begin{figure*}[t]
    \centering
    \includegraphics[width=0.8\textwidth]
    {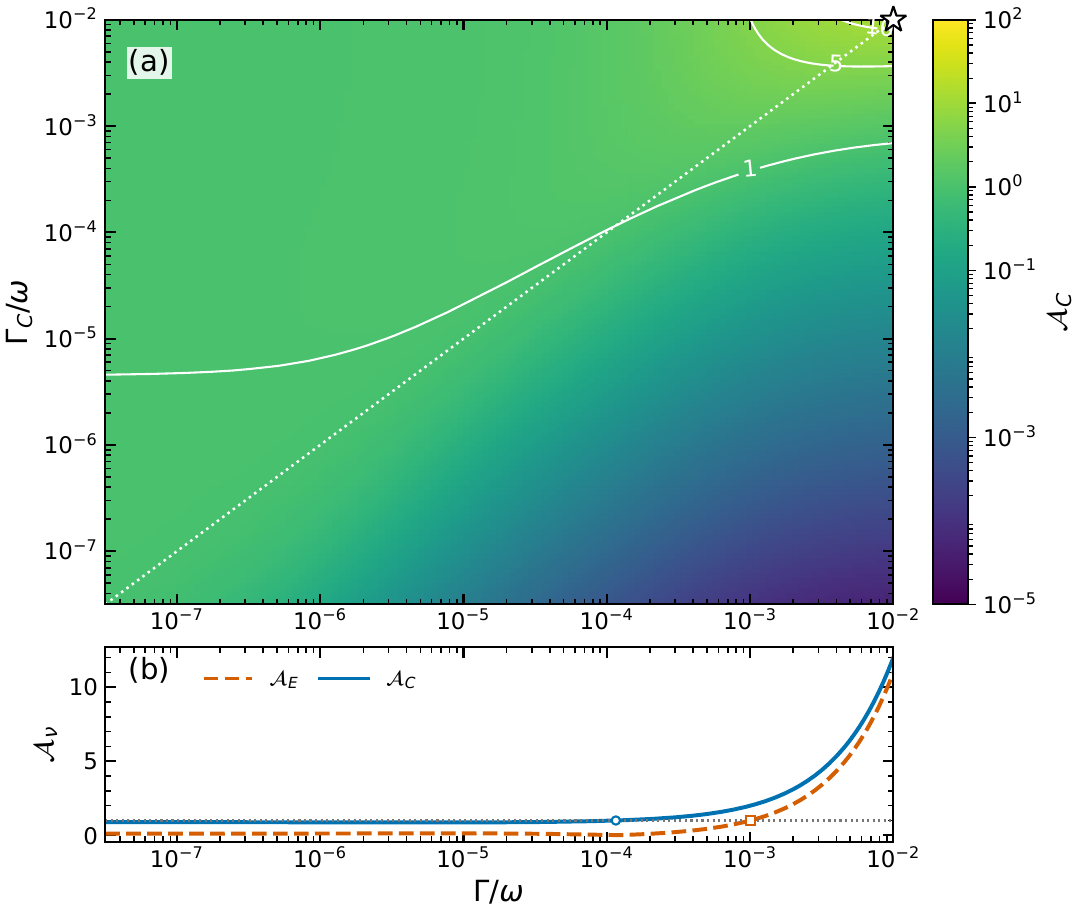}
    \caption{
    Thermal amplification with two nonlinear transport contacts.
    (a) Collector amplification \(\mathcal A_C\) versus
    \(\Gamma_E/\omega\) (horizontal axis) and
    \(\Gamma_C/\omega\) (vertical axis).
    The color scale is logarithmic, and white contours mark the
    labeled amplification levels. The dotted diagonal denotes
    \(\Gamma_E=\Gamma_C\), and the white star marks
    \(\Gamma_E=\Gamma_C=10^{-2}\omega\).
    (b) Amplification factors \(\mathcal A_E\) (orange dashed)
    and \(\mathcal A_C\) (blue solid) along that diagonal,
    with \(\Gamma_E=\Gamma_C=\Gamma\).
    Both panels use \(\gamma_E=\gamma_C=0\),
    \(\gamma_B/\omega=10^{-4}\), \(T_E/\omega=1.4\),
    \(T_C/\omega=0.1\), and \(T_B/\omega=0.75\),
    with \(\hbar=k_B=1\).
    The vanishing one-quantum transport rates can be achieved using the
    bath-spectrum-filtering discussed in
    Sec.~\ref{sec:experimental_realization}, in which the first
    sidebands are suppressed while the second sidebands are retained.
    }
    \label{fig:two_contact_landscape}
\end{figure*}
% -----------------------------------------------------------------------------

Substituting the dissipators in Eqs.~\eqref{eq:linear_dissipator} and \eqref{eq:nonlinear_dissipator} into the stationary heat-current
definition in Eq.~\eqref{eq:heat_current_definition} gives
\begin{align}
J_{\nu,\mathrm{ss}}
={}&
\omega\gamma_\nu
\left(\bar n_\nu-n_{\mathrm{ss}}\right)
\nonumber\\
&+
2\omega\Gamma_\nu
\left[
-M_{2,\mathrm{ss}}
+4\bar m_\nu n_{\mathrm{ss}}
+2\bar m_\nu
\right],
\nonumber\\
J_{B,\mathrm{ss}}
={}&
\omega\gamma_B
\left(\bar n_B-n_{\mathrm{ss}}\right).
\label{eq:two_contact_heat_currents}
\end{align}
Both transport currents depend on the stationary occupation
\(n_{\mathrm{ss}}\) and second factorial moment
\(M_{2,\mathrm{ss}}\), whereas the base current depends only
on the occupation. As in
Sec.~\ref{subsubsec:nonlinear_emitter_exact_equations},
the occupation equation involves the second factorial moment,
whose equation introduces higher moments. The nonlinear moment
hierarchy therefore does not provide a closed finite set of
equations for evaluating these currents.

We determine the required moments directly from the stationary
populations. Projecting Eq.~\eqref{eq:general_master_equation}
onto the oscillator Fock basis gives
\begin{equation}
\begin{aligned}
\frac{dp_j(t)}{dt}
={}&
(S+\gamma_\Sigma)
\bigl[(j+1)p_{j+1}-jp_j\bigr]
\\
&+
S\bigl[jp_{j-1}-(j+1)p_j\bigr]
\\
&+
\sum_{\nu=E,C}\Gamma_\nu
\Bigl\{
(\bar m_\nu+1)
\bigl[(j+2)(j+1)p_{j+2}
\\
&\qquad
-j(j-1)p_j\bigr]
+
\bar m_\nu
\bigl[j(j-1)p_{j-2}
\\
&\qquad
-(j+1)(j+2)p_j\bigr]
\Bigr\}.
\end{aligned}
\label{eq:two_contact_population_equation}
\end{equation}
Setting the left-hand side to zero determines the stationary populations
\(p_{j,\mathrm{ss}}\), subject to
\(p_{-1,\mathrm{ss}}=p_{-2,\mathrm{ss}}=0\) and
\(\sum_{j=0}^{\infty}p_{j,\mathrm{ss}}=1\).
We solve this stationary equation in a truncated Fock space and
increase the cutoff until the currents and differential responses
are converged. Differentiating the stationary population equation
and its normalization condition with respect to \(T_B\) gives
the temperature sensitivities needed to evaluate the current
derivatives. We then calculate the amplification numerically
using Eqs.~\eqref{eq:signed_thermal_gain} and
\eqref{eq:thermal_amplification_factor}. This approach avoids
moment closure and a perturbative expansion in the nonlinear
rates, allowing two-quantum coupling rates large relative to
the linear relaxation scale.

Figure~\ref{fig:two_contact_landscape}(a) shows the collector
amplification as the emitter and collector two-quantum rates
vary independently. The displayed landscape contains both
amplifying and nonamplifying regions, with strong amplification
when both nonlinear rates are sufficiently large. The equal-rate
diagonal passes through these regions and provides a representative
path for examining the onset and growth of amplification.
Figure~\ref{fig:two_contact_landscape}(b) resolves this diagonal
using the same parameters. For small positive nonlinear rates,
neither terminal amplifies. As the rates increase, the collector
amplification crosses unity before the emitter amplification.
These results establish strong amplification with two nonlinear
transport contacts in the harmonic working medium. We next examine how two-quantum exchange modifies the stationary
current responses to identify the mechanism responsible for
this amplification.

% .............................................................................
\subsubsection{Differential-response mechanism}
\label{subsubsec:nonlinear_emitter_collector_mechanism}
% .............................................................................

Differentiating the stationary currents in Eq.~\eqref{eq:two_contact_heat_currents} with respect to \(T_B\) gives
\begin{align}
\kappa_\nu
={}&-\omega\gamma_\nu\partial_{T_B}n_{\mathrm{ss}}
\nonumber\\
&+2\omega\Gamma_\nu
\left[-\partial_{T_B}M_{2,\mathrm{ss}}
+4\bar m_\nu\partial_{T_B}n_{\mathrm{ss}}\right],
\nonumber\\
\kappa_B
={}&\omega\gamma_B
\left[\bar n_B'-\partial_{T_B}n_{\mathrm{ss}}\right],
\label{eq:two_contact_current_responses}
\end{align}
where \(\bar n_B'\) is the base-reservoir occupation derivative
defined in Eq.~\eqref{eq:base_occupation_derivative}.
Because the transport-reservoir temperatures and all coupling
rates remain fixed, the emitter and collector currents respond
only through changes in \(n_{\mathrm{ss}}\) and
\(M_{2,\mathrm{ss}}\). In Eq.~\eqref{eq:two_contact_current_responses}, the base response \(\kappa_B\) also contains the explicit
temperature dependence of \(\bar n_B\), offset by the oscillator
occupation response.

Heat current conservation fixes the sum of the transport responses,
while their difference determines how the response is distributed
between the two terminals. Taking the emitter response minus the
collector response in Eq.~\eqref{eq:two_contact_current_responses}
gives
\begin{align}
\kappa_E-\kappa_C
={}&\omega\Bigl[\gamma_C-\gamma_E
+8\bigl(\Gamma_E\bar m_E-\Gamma_C\bar m_C\bigr)\Bigr]
\nonumber\\
&\times\partial_{T_B}n_{\mathrm{ss}}
+2\omega(\Gamma_C-\Gamma_E)
\partial_{T_B}M_{2,\mathrm{ss}}.
\label{eq:two_contact_response_contrast_general}
\end{align}
The second-factorial-moment contribution cancels from the
transport-response difference when \(\Gamma_E=\Gamma_C=\Gamma\),
where \(\Gamma\) is the common nonlinear rate. Substituting
this condition into
Eq.~\eqref{eq:two_contact_response_contrast_general} yields
\begin{equation}
\begin{aligned}
\kappa_E-\kappa_C
={}&\omega\left[\gamma_C-\gamma_E
+8\Gamma(\bar m_E-\bar m_C)\right]
\partial_{T_B}n_{\mathrm{ss}}.
\end{aligned}
\label{eq:two_contact_response_contrast_equal}
\end{equation}
The transport-response difference is thus proportional to the
oscillator occupation sensitivity, which remains determined by
the full stationary population dynamics. Combining
Eq.~\eqref{eq:two_contact_response_contrast_equal} with
\(\kappa_E+\kappa_C=-\kappa_B\) from
Eq.~\eqref{eq:differential_current_conservation} determines
both transport responses. When the magnitude of their difference
greatly exceeds \(|\kappa_B|\), the two responses have opposite
signs and individually exceed the base response in magnitude,
while their sum remains fixed at \(-\kappa_B\). We therefore
compare the transport-response difference with the base response
to quantify the resulting thermal amplification.

Dividing Eq.~\eqref{eq:two_contact_response_contrast_equal}
by the base response in
Eq.~\eqref{eq:two_contact_current_responses} gives
\begin{equation}
R_\kappa
=\frac{\kappa_E-\kappa_C}{\kappa_B}
=\frac{\Lambda\chi}{1-\chi}.
\label{eq:two_contact_response_contrast_ratio}
\end{equation}
Here, the dimensionless factors \(\Lambda\) and \(\chi\) are given by
\begin{align}
\Lambda
&=\frac{\gamma_C-\gamma_E
+8\Gamma(\bar m_E-\bar m_C)}{\gamma_B},
\label{eq:two_contact_transport_asymmetry}
\\
\chi
&=\frac{\partial_{T_B}n_{\mathrm{ss}}}{\bar n_B'}.
\label{eq:two_contact_occupation_susceptibility}
\end{align}
The coefficient \(\Lambda\) measures how strongly that occupation change produces a difference between the
transport currents, relative to the base coupling.  The susceptibility \(\chi\) measures how strongly the 
oscillator occupation follows a change in the base-reservoir occupation. To determine the amplification at each transport terminal, we use
\(R_\kappa=\beta_E-\beta_C\), which follows from the signed-gain
definition in Eq.~\eqref{eq:signed_thermal_gain}.
Combining this difference with
\(\beta_E+\beta_C=-1\) from
Eq.~\eqref{eq:signed_gain_conservation} gives
\begin{equation}
\beta_E=\frac{R_\kappa-1}{2},
\qquad
\beta_C=-\frac{R_\kappa+1}{2}.
\label{eq:two_contact_equal_rate_gains}
\end{equation}
On the positive-\(R_\kappa\) branch considered here, the collector gain has magnitude \((R_\kappa+1)/2\), so collector amplification
occurs for \(R_\kappa>1\). The emitter gain exceeds unity for \(R_\kappa>3\), beyond which both terminals amplify. Amplification is therefore governed by the ratio of the
transport-response difference \(\Lambda\chi\) to the base response \(1-\chi\).

Equation~\eqref{eq:two_contact_response_contrast_ratio} relates amplification to the oscillator occupation response and its
effect on transport. We now examine how the base temperature produces this response by redistributing the oscillator
populations, and how that redistribution controls two-quantum heat exchange.
% -----------------------------------------------------------------------------
\begin{figure*}[t]
    \centering
    \includegraphics[width=0.8\textwidth]
    {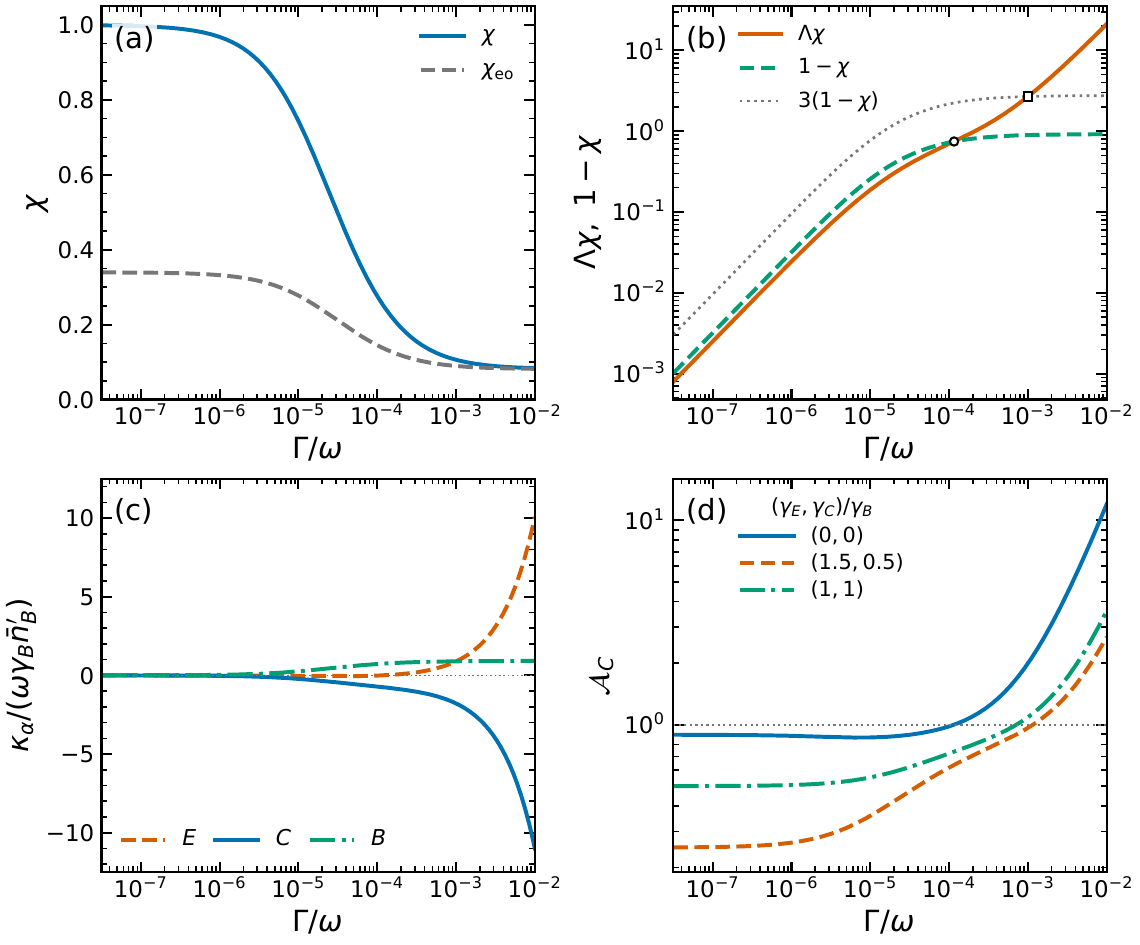}
    \caption{
    Population control, differential amplification, and the effect
    of single-quantum transport exchange along
    \(\Gamma_E=\Gamma_C=\Gamma\).
    All panels use \(\gamma_B/\omega=10^{-4}\),
    \(T_E/\omega=1.4\), \(T_C/\omega=0.1\), and
    \(T_B/\omega=0.75\), with \(\hbar=k_B=1\).
    Panels (a)--(c) use \(\gamma_E=\gamma_C=0\).
    Suppression of single-quantum transport coupling through
    bath spectral filtering is discussed in
    Sec.~\ref{sec:experimental_realization}.
    (a) Total occupation susceptibility \(\chi\) (blue solid)
    and the contribution \(\chi_{\mathrm{eo}}\) from redistribution
    between even and odd sectors (gray dashed), defined in
    Eqs.~\eqref{eq:two_contact_occupation_susceptibility} and
    \eqref{eq:two_contact_sector_weight_response}, respectively.
    (b) Normalized transport-response difference \(\Lambda\chi\)
    (orange solid), normalized base response \(1-\chi\)
    (green dashed), and \(3(1-\chi)\) (gray dotted).
    The open circle and square mark the collector and emitter
    amplification boundaries, respectively.
    (c) Signed total responses
    \(\kappa_\alpha/(\omega\gamma_B\bar n_B')\), with the
    emitter, collector, and base shown by orange dashed, blue solid,
    and green dash-dotted curves, respectively.
    The dotted horizontal line marks zero.
    (d) Collector amplification for
    \((\gamma_E,\gamma_C)/\gamma_B=(0,0)\) (blue solid),
    \((1.5,0.5)\) (orange dashed), and \((1,1)\)
    (green dash-dotted). The two nonzero cases have the same
    total single-quantum transport coupling,
    \(\gamma_E+\gamma_C=2\gamma_B\).
    The dotted horizontal line marks \(\mathcal A_C=1\).
    Each curve uses its own stationary state and temperature derivatives.
    }
    \label{fig:two_contact_mechanism}
\end{figure*}
% -----------------------------------------------------------------------------
In Fig.~\ref{fig:two_contact_mechanism}(a)--(c), we set
\(\gamma_E=\gamma_C=0\), suppressing single-quantum transport
exchange through bath spectral filtering as discussed in
Sec.~\ref{sec:experimental_realization}.
In this case, the transport reservoirs exchange only pairs of
quanta with the oscillator, connecting levels whose excitation
numbers differ by two in
Eq.~\eqref{eq:two_contact_population_equation}.
These transitions preserve the even and odd sectors, comprising
the even- and odd-numbered Fock levels, respectively.
Only the base-induced single-quantum transitions transfer
probability between the two sectors.
When pair exchange dominates, the transport reservoirs rapidly
redistribute populations within each sector, while the base
continues to influence their relative probabilities.
The base can therefore affect the stationary occupation even
when the nonlinear rates greatly exceed its coupling rate.
To identify the population changes responsible for this
sensitivity, we separate redistribution between the sectors
from changes within them.

Separating the stationary mean occupation into its even and
odd contributions gives
\[
n_{\mathrm{ss}}
=
P_{\mathrm{e},\mathrm{ss}}n_{\mathrm{e},\mathrm{ss}}
+
P_{\mathrm{o},\mathrm{ss}}n_{\mathrm{o},\mathrm{ss}}.
\]
Here,
\(P_{\mathrm{e},\mathrm{ss}}
=\sum_{j=0}^{\infty}p_{2j,\mathrm{ss}}\)
and
\(P_{\mathrm{o},\mathrm{ss}}
=\sum_{j=0}^{\infty}p_{2j+1,\mathrm{ss}}\)
are the probabilities of occupying the even and odd sectors,
respectively. The conditional mean occupations
\(n_{\mathrm{e},\mathrm{ss}}\) and
\(n_{\mathrm{o},\mathrm{ss}}\) are the mean excitation numbers
within the respective sectors.
Differentiating this weighted average with respect to \(T_B\)
and using
\(\partial_{T_B}P_{\mathrm{e},\mathrm{ss}}
=-\partial_{T_B}P_{\mathrm{o},\mathrm{ss}}\),
which follows from normalization, gives
\[
\begin{aligned}
\partial_{T_B}n_{\mathrm{ss}}
={}&
\left(
n_{\mathrm{o},\mathrm{ss}}-n_{\mathrm{e},\mathrm{ss}}
\right)
\partial_{T_B}P_{\mathrm{o},\mathrm{ss}}
\\
&+
P_{\mathrm{e},\mathrm{ss}}
\partial_{T_B}n_{\mathrm{e},\mathrm{ss}}
+
P_{\mathrm{o},\mathrm{ss}}
\partial_{T_B}n_{\mathrm{o},\mathrm{ss}}.
\end{aligned}
\]
The first term describes the occupation change caused by
redistribution between the sectors. The remaining terms
describe changes in the mean occupation within each sector.
Dividing the first term by \(\bar n_B'\), using the same
normalization as \(\chi\) in
Eq.~\eqref{eq:two_contact_occupation_susceptibility}, defines
\begin{equation}
\chi_{\mathrm{eo}}
=
\left(
n_{\mathrm{o},\mathrm{ss}}-n_{\mathrm{e},\mathrm{ss}}
\right)
\frac{\partial_{T_B}P_{\mathrm{o},\mathrm{ss}}}{\bar n_B'}.
\label{eq:two_contact_sector_weight_response}
\end{equation}
The quantity \(\chi_{\mathrm{eo}}\) is the contribution to
the occupation susceptibility \(\chi\) from changes in the relative
sector probabilities. The remainder,
\(\chi-\chi_{\mathrm{eo}}\), comes from changes within
the sectors. 

Figure~\ref{fig:two_contact_mechanism}(a) shows that the total
susceptibility \(\chi\) decreases with increasing nonlinear rate,
while \(\chi_{\mathrm{eo}}\) approaches the total response
in the regime of strong pair exchange.
At \(\Gamma/\omega=10^{-2}\), redistribution between the sectors
accounts for approximately \(98\%\) of the occupation
susceptibility. The oscillator therefore retains sensitivity
to the base temperature primarily through changes in the
relative probabilities of the even and odd sectors at this
operating point. Changes within the sectors also contribute
near the amplification onset.

This redistribution affects heat transport because the
two-quantum transition rates depend on the occupied Fock level
through the factors \((j+1)(j+2)\) and \(j(j-1)\) in
Eq.~\eqref{eq:two_contact_population_equation}.
For example, the upward transitions \(0\rightarrow2\) and
\(1\rightarrow3\) have matrix-element factors of two and six,
respectively. Changing the population distribution therefore
changes the rates of pair exchange, even though the transport
temperatures and coupling strengths remain fixed.
Equation~\eqref{eq:two_contact_response_contrast_equal} connects
this population response to the transport-response difference.
For \(\gamma_E=\gamma_C=0\), the coefficient multiplying
\(\partial_{T_B}n_{\mathrm{ss}}\) is
\(8\omega\Gamma(\bar m_E-\bar m_C)\).
The population analysis identifies how the base produces the
occupation response, while this coefficient determines how
strongly that response changes transport.
The unequal reservoir occupations provide the transport bias,
and increasing \(\Gamma\) strengthens the effect of an
occupation change on the transport-current difference.
This strengthening can outweigh the decrease in \(\chi\).

To determine when this population control produces amplification,
we compare the transport-response difference with the base response.
Normalizing Eq.~\eqref{eq:two_contact_response_contrast_equal}
and the base response in
Eq.~\eqref{eq:two_contact_current_responses} by
\(\omega\gamma_B\bar n_B'\) gives
\begin{align}
\frac{\kappa_E-\kappa_C}{\omega\gamma_B\bar n_B'}
&=\Lambda\chi,
\label{eq:two_contact_normalized_response_contrast}
\\
\frac{\kappa_B}{\omega\gamma_B\bar n_B'}
&=1-\chi.
\label{eq:two_contact_base_response_susceptibility}
\end{align}
These expressions place both responses on the same scale,
so their ratio directly determines \(R_\kappa\).
Figure~\ref{fig:two_contact_mechanism}(b) shows that
\(\Lambda\chi\) increases with \(\Gamma\) despite the decreasing
occupation susceptibility. When \(\Lambda\chi\) exceeds
\(1-\chi\), \(R_\kappa>1\) and the collector amplifies.
Its subsequent crossing of \(3(1-\chi)\) marks the emitter
amplification boundary. Beyond this crossing, \(R_\kappa>3\)
and both terminals amplify, as follows from
Eq.~\eqref{eq:two_contact_equal_rate_gains}.

Figure~\ref{fig:two_contact_mechanism}(c) shows how this amplification
appears in the individual current responses. Beyond the collector
amplification boundary, \(\kappa_E>0\) and \(\kappa_C<0\).
Because heat enters the oscillator from the emitter and leaves
through the collector in this regime, increasing \(T_B\) increases
both the heat supplied by the emitter and the heat delivered
to the collector. As amplification grows, these opposite changes
become large compared with the base-current change, while their
sum remains fixed at \(-\kappa_B\) by
Eq.~\eqref{eq:differential_current_conservation}.
The normalized base response remains finite in the regime of
strong amplification. The base thus controls a larger variation
in heat flow driven by the emitter--collector temperature bias
through its influence on the oscillator populations.

Figure~\ref{fig:two_contact_mechanism}(d) shows that restoring
single-quantum transport exchange reduces collector amplification
and shifts its onset to larger nonlinear rates in the matched
configurations. These additional channels connect the even and
odd sectors, reducing the sensitivity of their populations to
the base temperature. Equal single-quantum transport rates leave
\(\Lambda\) unchanged from its value at
\(\gamma_E=\gamma_C=0\), but reduce \(\chi\), thereby lowering
\(R_\kappa\) through
Eq.~\eqref{eq:two_contact_response_contrast_ratio}.
For the unequal rates shown, both \(\Lambda\) and \(\chi\)
decrease. The suppression therefore involves the modified
stationary population response and cannot be inferred from
the explicit single-quantum current terms alone.
Figure~\ref{fig:two_contact_mechanism} shows how base-induced
population redistribution controls two-quantum heat exchange,
producing large transport responses relative to the base response,
while additional single-quantum transport channels weaken this control.

% =============================================================================
\section{Experimental Feasibility}
\label{sec:experimental_realization}
% =============================================================================

The preceding analysis requires controllable one-quantum and two-quantum exchange while the working oscillator remains harmonic.
A route to these dissipative channels uses an auxiliary two-level system (TLS) coupled longitudinally to the oscillator.
The displacement-based construction and elimination of such an auxiliary were developed for nonlinear thermal transport in
Ref.~\cite{Naseem2026Rectification}. Here, we adapt this construction to a single TLS coupled to the
emitter and collector reservoirs, while the oscillator remains directly coupled to the base. We identify the spectral conditions
under which the eliminated transport channels acquire the thermal coefficients used in Sec.~\ref{sec:model_formalism}.

The oscillator--TLS Hamiltonian includes a longitudinal interaction~\cite{Didier2015, Touzard2019},
\begin{equation}
H_{\mathrm{comp}}
=
H_S
+
\frac{\omega_0}{2}\sigma_z
+
g\sigma_z(a+a^\dagger),
\label{eq:implementation_composite_hamiltonian}
\end{equation}
where \(H_S=\omega a^\dagger a\), \(\omega_0\) is the TLS transition frequency, and \(g\) is the
longitudinal coupling strength. The TLS operators satisfy \([\sigma_z,\sigma_\pm]=\pm2\sigma_\pm\), with
\(\sigma_x=\sigma_++\sigma_-\). The reservoir interaction is
\[
H_{\mathrm{comp},R}
=
\sigma_x(Y_E+Y_C)+(a+a^\dagger)Y_B,
\]
where
\(Y_\alpha=\sum_k (h_{\alpha k}b_{\alpha k} +h_{\alpha k}^*b_{\alpha k}^\dagger)\)
is a Hermitian coupling operator of reservoir \(\alpha\). The reservoir Hamiltonians and thermal states are those introduced
in Sec.~\ref{subsec:microscopic_model_master_equation}. The three reservoirs are independent, and their coupling operators
have zero thermal means.

To expose the oscillator transitions induced by a TLS flip, we apply the conditional displacement
\begin{equation}
U
=
\exp\!\left[
\frac{\xi}{2}\sigma_z(a^\dagger-a)
\right],
\qquad
\xi=\frac{2g}{\omega}.
\label{eq:implementation_displacement}
\end{equation}
Here, \(\xi\) is the dimensionless displacement associated with a TLS transition. Transforming the composite state according to
\(\widetilde\varrho=U\varrho U^\dagger\) gives the Hamiltonian
\[
\widetilde H_{\mathrm{comp}}
=
UH_{\mathrm{comp}}U^\dagger
=
H_S+\frac{\omega_0}{2}\sigma_z-\frac{g^2}{\omega}.
\]
Discarding the constant shift, we write \(\widetilde H_0=H_S+\omega_0\sigma_z/2\).
We retain \(a\) and \(\sigma_\pm\) for the canonical oscillator and TLS operators in this representation.
The transformed TLS lowering operator is \(U\sigma_-U^\dagger=\sigma_-D(-\xi)\), where
\(D(-\xi)=\exp[-\xi(a^\dagger-a)]\). Expanding the displacement gives
\begin{equation}
\begin{aligned}
D(-\xi)
={}&
1-\xi^2\left(N+\frac12\right)
+\xi a-\xi a^\dagger
\\
&+
\frac{\xi^2}{2}
\left(a^2+a^{\dagger2}\right)
+O(\xi^3),
\end{aligned}
\label{eq:implementation_sideband_expansion}
\end{equation}
where \(N=a^\dagger a\). The expansion requires \(\xi\sqrt{n_{\max}+1}\ll1\), where \(n_{\max}\) denotes
the upper end of the appreciably populated Fock levels. The number-diagonal factor
\(F_0=1-\xi^2(N+1/2)\) generates the carrier transition without changing the oscillator excitation number.
The terms proportional to \(a\) and \(a^\dagger\) generate one-quantum sidebands at \(\omega_0\pm\omega\), while
\(a^2\) and \(a^{\dagger2}\) generate two-quantum sidebands at \(\omega_0\pm2\omega\).
We take \(\omega_0>2\omega\), so these frequencies are positive. Their leading dissipative coefficients carry factors
\(\xi^2\) and \(\xi^4/4\), respectively.

Let \(G_\alpha(\Omega)\) denote the two-sided correlation spectrum of reservoir \(\alpha\),
\[
G_\alpha(\Omega)
=
\int_{-\infty}^{\infty}dt\,
e^{i\Omega t}
\langle Y_\alpha(t)Y_\alpha(0)\rangle.
\]
Thermal equilibrium implies
\begin{equation}
G_\alpha(-\Omega)
=
e^{-\Omega/T_\alpha}G_\alpha(\Omega),
\qquad \Omega>0.
\label{eq:implementation_reservoir_detailed_balance}
\end{equation}
Under the Born--Markov and secular approximations, the resolved components of Eq.~\eqref{eq:implementation_sideband_expansion}
generate separate dissipators with coefficients sampled at their respective Bohr frequencies.
We require reservoir correlation times to be short compared with the relevant dissipative timescales, and dissipative linewidths
to be small compared with the separations between the retained frequency components.
Following the frequency-resolved construction of Ref.~\cite{Naseem2026Rectification}, the composite master equation is
\begin{equation}
\begin{aligned}
\frac{d\widetilde\varrho(t)}{dt}
={}&
-i[\widetilde H_0,\widetilde\varrho(t)]
+
\sum_{\nu=E,C}
\mathcal K_\nu\widetilde\varrho(t)
+
\mathcal K_B\widetilde\varrho(t).
\end{aligned}
\label{eq:implementation_composite_master_equation}
\end{equation}
Here, \(\mathcal K_\nu\) and \(\mathcal K_B\) act on the joint TLS--oscillator state.
As in Sec.~\ref{subsec:microscopic_model_master_equation}, reservoir-induced Lamb-shift terms are omitted from the
displayed dynamics. Using the carrier factor \(F_0\) and the leading contribution of each retained sideband gives
\begin{align}
\mathcal K_\nu
={}&
G_\nu(\omega_0)\mathcal D[\sigma_-F_0]
+
G_\nu(-\omega_0)\mathcal D[\sigma_+F_0]
\nonumber\\
&+
\sum_{q=1}^{2}
\frac{\xi^{2q}}{(q!)^2}
\Bigl\{
G_\nu(\omega_0+q\omega)
\mathcal D[\sigma_-a^q]
\nonumber\\
&\qquad+
G_\nu(-\omega_0-q\omega)
\mathcal D[\sigma_+a^{\dagger q}]
\nonumber\\
&\qquad+
G_\nu(\omega_0-q\omega)
\mathcal D[\sigma_-a^{\dagger q}]
\nonumber\\
&\qquad+
G_\nu(-\omega_0+q\omega)
\mathcal D[\sigma_+a^q]
\Bigr\}.
\label{eq:implementation_composite_transport_generators}
\end{align}
The index \(q=1,2\) denotes the number of oscillator quanta exchanged in a sideband transition.
For example, \(\sigma_-a^q\) removes \(q\) oscillator quanta while relaxing the TLS, whereas \(\sigma_+a^q\) removes the
same number while exciting the TLS. Both pathways therefore contribute to oscillator relaxation,
although they sample different reservoir frequencies. The base coupling transforms as
\(U(a+a^\dagger)U^\dagger=a+a^\dagger-\xi\sigma_z\). Its frequency-resolved generator is consequently
\begin{equation}
\begin{aligned}
\mathcal K_B
={}&
\gamma_B(\bar n_B+1)\mathcal D[a]
+
\gamma_B\bar n_B\mathcal D[a^\dagger]
+
\xi^2G_B(0)\mathcal D[\sigma_z],
\end{aligned}
\label{eq:implementation_composite_base_generator}
\end{equation}
where
\(G_B(\omega)=\gamma_B(\bar n_B+1)\) and
\(G_B(-\omega)=\gamma_B\bar n_B\).
The zero-frequency term produces TLS dephasing and leaves the TLS populations unchanged.
The oscillator therefore retains the linear base contact specified in Eq.~\eqref{eq:general_master_equation}.

To eliminate the auxiliary, we take its bare carrier relaxation as the fast reference dynamics and retain the
displacement-dependent carrier corrections and sidebands as perturbations.
The carrier population equation is
\[
\frac{dp_e(t)}{dt}
=
\sum_{\nu=E,C}
\left[
G_\nu(-\omega_0)(1-p_e(t))
-
G_\nu(\omega_0)p_e(t)
\right].
\]
Setting its left-hand side to zero gives
\begin{equation}
p_e
=
\frac{
G_E(-\omega_0)+G_C(-\omega_0)
}{
\displaystyle
\sum_{\nu=E,C}
\left[G_\nu(\omega_0)+G_\nu(-\omega_0)\right]
},
\qquad
p_g=1-p_e.
\label{eq:implementation_shared_tls_population}
\end{equation}
Here, \(p_e\) and \(p_g\) are the stationary populations of the fast carrier reference state.
Both transport reservoirs determine this single population distribution. It is therefore generally thermal at neither
\(T_E\) nor \(T_C\). Its relaxation rate is \(\gamma_0=\sum_{\nu=E,C} [G_\nu(\omega_0)+G_\nu(-\omega_0)]\).
We require the sideband and base-induced transition rates, including their Fock-state matrix elements over the populated
levels, to remain small compared with \(\gamma_0\).

After the fast TLS relaxation, the leading eliminated state is \(\widetilde\varrho(t)\simeq\rho(t)\otimes\tau\), where
\(\tau=p_e|e\rangle\langle e|+p_g|g\rangle\langle g|\) and \(\rho(t)=\operatorname{Tr}_{\mathrm{TLS}}\widetilde\varrho(t)\)
is the reduced oscillator state in the displaced representation. For any oscillator operator \(X\), the partial-trace identities
are
\[
\begin{aligned}
\operatorname{Tr}_{\mathrm{TLS}}
\mathcal D[\sigma_-X](\rho\otimes\tau)
&=
p_e\mathcal D[X]\rho,
\\
\operatorname{Tr}_{\mathrm{TLS}}
\mathcal D[\sigma_+X](\rho\otimes\tau)
&=
p_g\mathcal D[X]\rho.
\end{aligned}
\]
Applying these identities to the second sidebands in Eq.~\eqref{eq:implementation_composite_transport_generators} yields
\begin{equation}
\begin{aligned}
\widetilde\Gamma_\nu^-
={}&
\frac{\xi^4}{4}
\Bigl[
p_eG_\nu(\omega_0+2\omega)
+
p_gG_\nu(-\omega_0+2\omega)
\Bigr],
\\
\widetilde\Gamma_\nu^+
={}&
\frac{\xi^4}{4}
\Bigl[
p_eG_\nu(\omega_0-2\omega)
+
p_gG_\nu(-\omega_0-2\omega)
\Bigr].
\end{aligned}
\label{eq:implementation_effective_two_quantum_rates}
\end{equation}
The coefficients \(\widetilde\Gamma_\nu^-\) and \(\widetilde\Gamma_\nu^+\) describe removal and addition of
two oscillator quanta, respectively. Each contains a TLS-relaxation pathway weighted by \(p_e\)
and a TLS-excitation pathway weighted by \(p_g\). The one-quantum coefficients \(\widetilde\gamma_\nu^\pm\) follow from
Eq.~\eqref{eq:implementation_effective_two_quantum_rates} by replacing \(2\omega\) with \(\omega\) and \(\xi^4/4\)
with \(\xi^2\).

Using this factorization and taking the TLS trace of Eq.~\eqref{eq:implementation_composite_master_equation} gives the leading oscillator master equation
\begin{equation}
\begin{aligned}
\frac{d\rho(t)}{dt}
={}&
-i[H_S,\rho(t)]
\\
&+
\gamma_B(\bar n_B+1)\mathcal D[a]\rho(t)
+
\gamma_B\bar n_B\mathcal D[a^\dagger]\rho(t)
\\
&+
\sum_{\nu=E,C}
\Bigl[
\widetilde\gamma_\nu^-\mathcal D[a]
+
\widetilde\gamma_\nu^+\mathcal D[a^\dagger]
\\
&\qquad\qquad
+
\widetilde\Gamma_\nu^-\mathcal D[a^2]
+
\widetilde\Gamma_\nu^+\mathcal D[a^{\dagger2}]
\Bigr]\rho(t)
\\
&+
\gamma_\phi\mathcal D[N]\rho(t).
\end{aligned}
\label{eq:implementation_reduced_oscillator_master_equation}
\end{equation}
The additional number-dephasing coefficient is
\begin{equation}
\gamma_\phi
=
\xi^4
\sum_{\nu=E,C}
\left[
p_eG_\nu(\omega_0)+p_gG_\nu(-\omega_0)
\right].
\label{eq:dephasing}
\end{equation}
It follows from the carrier factor through \(\mathcal D[F_0]=\xi^4\mathcal D[N]\).
Since \(\langle j|\mathcal D[N]\rho|j\rangle=0\) and \(\operatorname{Tr}[H_S\mathcal D[N]\rho]=0\),
this term leaves the oscillator population equations and harmonic-energy balance unchanged.
The TLS-dephasing term in Eq.~\eqref{eq:implementation_composite_base_generator}
vanishes upon taking the partial trace. Equation~\eqref{eq:implementation_reduced_oscillator_master_equation}
retains the leading contribution of each exchange channel and does not constitute a complete expansion through fourth order
in \(\xi\). When both sideband families are retained, omitted first-sideband
and higher-order elimination corrections must remain negligible on the scale of the two-quantum effects under consideration.
Suppressing the first sidebands removes the corrections associated with those frequencies, while the remaining approximation
conditions still apply.

The reduced equation contains the required oscillator jump operators. The shared auxiliary population makes a further
spectral condition necessary in general to recover the thermal coefficients of Sec.~\ref{sec:model_formalism}.
Using Eq.~\eqref{eq:implementation_reservoir_detailed_balance} in Eq.~\eqref{eq:implementation_effective_two_quantum_rates} gives
\begin{equation}
\begin{aligned}
&\widetilde\Gamma_\nu^+
-
e^{-2\omega/T_\nu}\widetilde\Gamma_\nu^-
\\
&\quad=
\frac{\xi^4}{4}
\left(p_e-p_ge^{-\omega_0/T_\nu}\right)
\\
&\qquad\times
\left[
G_\nu(\omega_0-2\omega)
-
e^{-2\omega/T_\nu}G_\nu(\omega_0+2\omega)
\right].
\end{aligned}
\label{eq:implementation_thermal_matching_identity}
\end{equation}
Thus, the required ratio can be recovered by controlling the relative second-sideband weights, independently of
whether the TLS population is thermal at \(T_\nu\). The same calculation applies to the first sidebands.
A sufficient condition for either retained channel is
\begin{equation}
G_\nu(\omega_0-q\omega)
=
e^{-q\omega/T_\nu}
G_\nu(\omega_0+q\omega),
\qquad q=1,2.
\label{eq:implementation_spectral_matching}
\end{equation}
Equation~\eqref{eq:implementation_spectral_matching} constrains the complete thermal correlation spectrum,
including its occupation factors. Under this condition, the eliminated coefficients satisfy
\[
\widetilde\gamma_\nu^+
=
e^{-\omega/T_\nu}\widetilde\gamma_\nu^-,
\qquad
\widetilde\Gamma_\nu^+
=
e^{-2\omega/T_\nu}\widetilde\Gamma_\nu^-.
\]
Identifying
\begin{equation}
\gamma_\nu
=
\widetilde\gamma_\nu^-
-
\widetilde\gamma_\nu^+,
\qquad
\Gamma_\nu
=
\widetilde\Gamma_\nu^-
-
\widetilde\Gamma_\nu^+,
\label{eq:implementation_rate_identification}
\end{equation}
therefore gives
\[
\begin{aligned}
\widetilde\gamma_\nu^-&=\gamma_\nu(\bar n_\nu+1),
&
\widetilde\gamma_\nu^+&=\gamma_\nu\bar n_\nu,
\\
\widetilde\Gamma_\nu^-&=\Gamma_\nu(\bar m_\nu+1),
&
\widetilde\Gamma_\nu^+&=\Gamma_\nu\bar m_\nu.
\end{aligned}
\]
Substituting these relations into Eq.~\eqref{eq:implementation_reduced_oscillator_master_equation}
recovers the dissipators in Eqs.~\eqref{eq:linear_dissipator} and \eqref{eq:nonlinear_dissipator}, together with the
number-dephasing term given in Eq.~\eqref{eq:dephasing}. 

Spectral filtering provides control over which sidebands contribute to these effective rates.
Reservoir interfaces containing resonant modes can suppress unwanted transition frequencies while retaining selected
spectral bands, as developed for quantum heat management and reservoir-assisted cooling in
Refs.~\cite{Naseem2020HeatManager,Naseem_2020,Naseem2021ReservoirCooling}. The selected bands must remain compatible with the reservoir
correlation-time and frequency-resolution conditions stated above. For predominantly two-quantum exchange, the first sidebands
are suppressed while the carrier and second sidebands remain active.
The ideal condition
\[
G_\nu(\omega_0-\omega)
=
G_\nu(\omega_0+\omega)
=0
\]
sets the leading one-quantum coefficients to zero. The retained second-sideband weights are then chosen to
satisfy Eq.~\eqref{eq:implementation_spectral_matching}. This gives the effective transport-channel structure used
in Fig.~\ref{fig:two_contact_landscape}(a) and Fig.~\ref{fig:two_contact_mechanism}(a)--(c). Finite filtering leaves residual one-quantum exchange,
whose strength follows from the eliminated first-sideband coefficients. Conversely, suppressing the second sidebands selects
predominantly one-quantum exchange, while retaining both families allows mixed coupling. Channel selection and the relative weighting required for
thermal matching are therefore distinct parts of the spectral design. The construction provides a proposed route to the required
dissipators under these conditions.

% =============================================================================
\section{Conclusions}
\label{sec:conclusions}
% =============================================================================

We have shown that nonlinear system-reservoir coupling enables thermal
amplification in a harmonic working medium. For the fully linear
benchmark, we recover the absence of amplification when the base
temperature varies at fixed transport temperatures and coupling rates.
Introducing two-quantum exchange at the emitter enables collector
amplification, with its onset predicted by a controlled weak-nonlinearity
expansion. Allowing two-quantum exchange at both transport contacts
produces strong amplification in the operating regimes examined.

The amplification mechanism combines base-induced population control
with nonlinear transport exchange. In the absence of single-quantum
transport coupling, two-quantum transitions preserve the even and odd
Fock sectors, while the base connects them. When two-quantum exchange
dominates, redistribution between these sectors accounts for most of
the remaining oscillator occupation response in the regime studied.
At equal nonlinear rates, the second-factorial-moment contribution
cancels from the difference between the transport responses.
The resulting identity connects this difference to the unequal
transport-reservoir occupations and the oscillator occupation response.
A large transport-response difference relative to the base response
allows strong amplification through oppositely signed emitter and
collector responses, whose sum is fixed by differential current
conservation. Restoring single-quantum transport exchange reduces
base control of the populations and suppresses amplification in the
matched configurations examined. The selected regimes also exhibit
a small stationary base current, a property distinct from the current
derivatives that determine amplification.

We have proposed a realization using an auxiliary two-level system
longitudinally coupled to the oscillator. Bath spectral filtering
selects the exchange channels, while appropriate relative spectral
weights recover their thermal coefficients. Eliminating the auxiliary
under the stated conditions yields the required leading oscillator
dynamics, together with number dephasing that leaves the population
dynamics unchanged. These results identify nonlinear reservoir
coupling as a resource for controlling and amplifying heat transport
without requiring intrinsic anharmonicity of the working medium.

% =============================================================================
\begin{acknowledgments}
M. Tahir Naseem acknowledges support from the National Center for Quantum Computing (NCQC), Pakistan.
\end{acknowledgments}
% =============================================================================

%===============================================================================
% Data Availability
%===============================================================================
\section*{Data Availability}
The numerical code used to generate the figures presented in this work is publicly available in Ref.~\cite{DataRepository}.
%===============================================================================

% =============================================================================
% Bibliography
% =============================================================================

\bibliography{qtt}

@article{Goold2016,
  author     = {Goold, John and Huber, Marcus and Riera, Arnau and del Rio, L{\'{i}}dia and Skrzypczyk, Paul},
  title      = {{The role of quantum information in thermodynamics---a topical review}},
  journal    = {Journal of Physics A: Mathematical and Theoretical},
  year       = {2016},
  volume     = {49},
  number     = {14},
  pages      = {143001},
  doi        = {10.1088/1751-8113/49/14/143001},
  url        = {https://doi.org/10.1088/1751-8113/49/14/143001},
}

@article{Vinjanampathy2016,
  author     = {Vinjanampathy, Sai and Anders, Janet},
  title      = {{Quantum thermodynamics}},
  journal    = {Contemporary Physics},
  year       = {2016},
  volume     = {57},
  number     = {4},
  pages      = {545--579},
  doi        = {10.1080/00107514.2016.1201896},
  url        = {https://doi.org/10.1080/00107514.2016.1201896},
}

@article{Benenti2017,
  author     = {Benenti, Giuliano and Casati, Giulio and Saito, Keiji and Whitney, Robert S.},
  title      = {{Fundamental aspects of steady-state conversion of heat to work at the nanoscale}},
  journal    = {Physics Reports},
  year       = {2017},
  volume     = {694},
  pages      = {1--124},
  doi        = {10.1016/j.physrep.2017.05.008},
  url        = {https://doi.org/10.1016/j.physrep.2017.05.008},
}

@article{SegalAgarwalla2016,
  author     = {Segal, Dvira and Agarwalla, Bijay Kumar},
  title      = {{Vibrational Heat Transport in Molecular Junctions}},
  journal    = {Annual Review of Physical Chemistry},
  year       = {2016},
  volume     = {67},
  number     = {1},
  pages      = {185--209},
  doi        = {10.1146/annurev-physchem-040215-112103},
  url        = {https://doi.org/10.1146/annurev-physchem-040215-112103},
}

@inproceedings{Naseem2024HeatCurrents,
  author     = {Naseem, M. Tahir and M{\"u}stecapl{\i}o{\u{g}}lu, {\"O}zg{\"u}r E.},
  editor     = {Aydiner, Ekrem and Sidharth, Burra G. and Michelini, Marisa and Corda, Christian},
  title      = {Fundamentals and Applications of Heat Currents in Quantum Systems},
  booktitle  = {Frontiers of Fundamental Physics FFP16},
  series     = {Springer Proceedings in Physics},
  year       = {2024},
  volume     = {392},
  pages      = {133--163},
  publisher  = {Springer},
  address    = {Cham},
  doi        = {10.1007/978-3-031-38477-6_8},
  url        = {https://doi.org/10.1007/978-3-031-38477-6_8},
  isbn       = {978-3-031-38477-6},
}

@book{BreuerPetruccione2002,
  author     = {Breuer, Heinz-Peter and Petruccione, Francesco},
  title      = {The Theory of Open Quantum Systems},
  year       = {2002},
  publisher  = {Oxford University Press},
  address    = {Oxford},
  isbn       = {9780198520634},
}

@article{Kosloff2013,
  author     = {Kosloff, Ronnie},
  title      = {Quantum Thermodynamics: A Dynamical Viewpoint},
  journal    = {Entropy},
  year       = {2013},
  volume     = {15},
  number     = {6},
  pages      = {2100--2128},
  doi        = {10.3390/e15062100},
  url        = {https://doi.org/10.3390/e15062100},
}

@article{Pekola2015,
  author     = {Pekola, Jukka P.},
  title      = {{Towards quantum thermodynamics in electronic circuits}},
  journal    = {Nature Physics},
  year       = {2015},
  volume     = {11},
  number     = {2},
  pages      = {118--123},
  doi        = {10.1038/nphys3169},
  url        = {https://doi.org/10.1038/nphys3169},
}

@article{PekolaKarimi2021,
  author     = {Pekola, Jukka P. and Karimi, Bayan},
  title      = {{Colloquium: Quantum heat transport in condensed matter systems}},
  journal    = {Reviews of Modern Physics},
  year       = {2021},
  volume     = {93},
  number     = {4},
  pages      = {041001},
  doi        = {10.1103/revmodphys.93.041001},
  url        = {https://doi.org/10.1103/revmodphys.93.041001},
}

@article{Rego1998,
  author     = {Rego, Luis G. C. and Kirczenow, George},
  title      = {{Quantized Thermal Conductance of Dielectric Quantum Wires}},
  journal    = {Physical Review Letters},
  year       = {1998},
  volume     = {81},
  number     = {1},
  pages      = {232--235},
  doi        = {10.1103/physrevlett.81.232},
  url        = {https://doi.org/10.1103/physrevlett.81.232},
}

@article{Schwab2000,
  author     = {Schwab, K. and Henriksen, E. A. and Worlock, J. M. and Roukes, M. L.},
  title      = {{Measurement of the quantum of thermal conductance}},
  journal    = {Nature},
  year       = {2000},
  volume     = {404},
  number     = {6781},
  pages      = {974--977},
  doi        = {10.1038/35010065},
  url        = {https://doi.org/10.1038/35010065},
}

@article{Jezouin2013,
  author     = {Jezouin, S. and Parmentier, F. D. and Anthore, A. and Gennser, U. and Cavanna, A. and Jin, Y. and Pierre, F.},
  title      = {{Quantum Limit of Heat Flow Across a Single Electronic Channel}},
  journal    = {Science},
  year       = {2013},
  volume     = {342},
  number     = {6158},
  pages      = {601--604},
  doi        = {10.1126/science.1241912},
  url        = {https://doi.org/10.1126/science.1241912},
}

@article{Meschke2006,
  author     = {Meschke, Matthias and Guichard, Wiebke and Pekola, Jukka P.},
  title      = {{Single-mode heat conduction by photons}},
  journal    = {Nature},
  year       = {2006},
  volume     = {444},
  number     = {7116},
  pages      = {187--190},
  doi        = {10.1038/nature05276},
  url        = {https://doi.org/10.1038/nature05276},
}

@article{Partanen2016,
  author     = {Partanen, Matti and Tan, Kuan Yen and Govenius, Joonas and Lake, Russell E. and M{\"{a}}kel{\"{a}}, Miika K. and Tanttu, Tuomo and M{\"{o}}tt{\"{o}}nen, Mikko},
  title      = {{Quantum-limited heat conduction over macroscopic distances}},
  journal    = {Nature Physics},
  year       = {2016},
  volume     = {12},
  number     = {5},
  pages      = {460--464},
  doi        = {10.1038/nphys3642},
  url        = {https://doi.org/10.1038/nphys3642},
}

@article{Giazotto2012,
  author     = {Giazotto, Francesco and Mart{\'{i}}nez-P{\'{e}}rez, Mar{\'{i}}a Jos{\'{e}}},
  title      = {{The Josephson heat interferometer}},
  journal    = {Nature},
  year       = {2012},
  volume     = {492},
  number     = {7429},
  pages      = {401--405},
  doi        = {10.1038/nature11702},
  url        = {https://doi.org/10.1038/nature11702},
}

@article{Ronzani2018,
  author     = {Ronzani, Alberto and Karimi, Bayan and Senior, Jorden and Chang, Yu-Cheng and Peltonen, Joonas T. and Chen, ChiiDong and Pekola, Jukka P.},
  title      = {{Tunable photonic heat transport in a quantum heat valve}},
  journal    = {Nature Physics},
  year       = {2018},
  volume     = {14},
  number     = {10},
  pages      = {991--995},
  doi        = {10.1038/s41567-018-0199-4},
  url        = {https://doi.org/10.1038/s41567-018-0199-4},
}

@article{MartinezPerez2015,
  author     = {Mart{\'{i}}nez-P{\'{e}}rez, Maria Jos{\'{e}} and Fornieri, Antonio and Giazotto, Francesco},
  title      = {{Rectification of electronic heat current by a hybrid thermal diode}},
  journal    = {Nature Nanotechnology},
  year       = {2015},
  volume     = {10},
  number     = {4},
  pages      = {303--307},
  doi        = {10.1038/nnano.2015.11},
  url        = {https://doi.org/10.1038/nnano.2015.11},
}

@article{Senior2020,
  author     = {Senior, Jorden and Gubaydullin, Azat and Karimi, Bayan and Peltonen, Joonas T. and Ankerhold, Joachim and Pekola, Jukka P.},
  title      = {{Heat rectification via a superconducting artificial atom}},
  journal    = {Communications Physics},
  year       = {2020},
  volume     = {3},
  number     = {1},
  pages      = {40},
  doi        = {10.1038/s42005-020-0307-5},
  url        = {https://doi.org/10.1038/s42005-020-0307-5},
}

@article{Gubaydullin2022,
  author     = {Gubaydullin, Azat and Thomas, George and Golubev, Dmitry S. and Lvov, Dmitrii and Peltonen, Joonas T. and Pekola, Jukka P.},
  title      = {{Photonic heat transport in three terminal superconducting circuit}},
  journal    = {Nature Communications},
  year       = {2022},
  volume     = {13},
  number     = {1},
  pages      = {1552},
  doi        = {10.1038/s41467-022-29078-x},
  url        = {https://doi.org/10.1038/s41467-022-29078-x},
}

@article{Li2012,
  author     = {Li, Nianbei and Ren, Jie and Wang, Lei and Zhang, Gang and H{\"{a}}nggi, Peter and Li, Baowen},
  title      = {{Colloquium: Phononics: Manipulating heat flow with electronic analogs and beyond}},
  journal    = {Reviews of Modern Physics},
  year       = {2012},
  volume     = {84},
  number     = {3},
  pages      = {1045--1066},
  doi        = {10.1103/revmodphys.84.1045},
  url        = {https://doi.org/10.1103/revmodphys.84.1045},
}

@article{WangLi2007,
  author     = {Wang, Lei and Li, Baowen},
  title      = {{Thermal Logic Gates: Computation with Phonons}},
  journal    = {Physical Review Letters},
  year       = {2007},
  volume     = {99},
  number     = {17},
  pages      = {177208},
  doi        = {10.1103/physrevlett.99.177208},
  url        = {https://doi.org/10.1103/physrevlett.99.177208},
}

@article{Castelli2023,
  author     = {Castelli, Lorenzo and Zhu, Qing and Shimokusu, Trevor J. and Wehmeyer, Geoff},
  title      = {{A three-terminal magnetic thermal transistor}},
  journal    = {Nature Communications},
  year       = {2023},
  volume     = {14},
  number     = {1},
  pages      = {393},
  doi        = {10.1038/s41467-023-36056-4},
  url        = {https://doi.org/10.1038/s41467-023-36056-4},
}

@article{Saira2007,
  author     = {Saira, Olli-Pentti and Meschke, Matthias and Giazotto, Francesco and Savin, Alexander M. and M{\"{o}}tt{\"{o}}nen, Mikko and Pekola, Jukka P.},
  title      = {{Heat Transistor: Demonstration of Gate-Controlled Electronic Refrigeration}},
  journal    = {Physical Review Letters},
  year       = {2007},
  volume     = {99},
  number     = {2},
  pages      = {027203},
  doi        = {10.1103/physrevlett.99.027203},
  url        = {https://doi.org/10.1103/physrevlett.99.027203},
}

@article{BenAbdallah2014,
  author     = {Ben-Abdallah, Philippe and Biehs, Svend-Age},
  title      = {{Near-Field Thermal Transistor}},
  journal    = {Physical Review Letters},
  year       = {2014},
  volume     = {112},
  number     = {4},
  pages      = {044301},
  doi        = {10.1103/physrevlett.112.044301},
  url        = {https://doi.org/10.1103/physrevlett.112.044301},
}

@article{Li2006,
  author     = {Li, Baowen and Wang, Lei and Casati, Giulio},
  title      = {{Negative differential thermal resistance and thermal transistor}},
  journal    = {Applied Physics Letters},
  year       = {2006},
  volume     = {88},
  number     = {14},
  pages      = {143501},
  doi        = {10.1063/1.2191730},
  url        = {https://doi.org/10.1063/1.2191730},
}

@article{Joulain2016,
  author     = {Joulain, Karl and Drevillon, J{\'{e}}r{\'{e}}mie and Ezzahri, Youn{\`{e}}s and Ordonez-Miranda, Jose},
  title      = {{Quantum Thermal Transistor}},
  journal    = {Physical Review Letters},
  year       = {2016},
  volume     = {116},
  number     = {20},
  pages      = {200601},
  doi        = {10.1103/physrevlett.116.200601},
  url        = {https://doi.org/10.1103/physrevlett.116.200601},
}

@article{Mandarino2021,
  author     = {Mandarino, Antonio and Joulain, Karl and Dom{\'{i}}nguez G{\'{o}}mez, Melisa and Bellomo, Bruno},
  title      = {{Thermal Transistor Effect in Quantum Systems}},
  journal    = {Physical Review Applied},
  year       = {2021},
  volume     = {16},
  number     = {3},
  pages      = {034026},
  doi        = {10.1103/physrevapplied.16.034026},
  url        = {https://doi.org/10.1103/physrevapplied.16.034026},
}

@article{Zhang2018,
  author     = {Zhang, Yanchao and Yang, Zhimin and Zhang, Xin and Lin, Bihong and Lin, Guoxing and Chen, Jincan},
  title      = {{Coulomb-coupled quantum-dot thermal transistors}},
  journal    = {EPL (Europhysics Letters)},
  year       = {2018},
  volume     = {122},
  number     = {1},
  pages      = {17002},
  doi        = {10.1209/0295-5075/122/17002},
  url        = {https://doi.org/10.1209/0295-5075/122/17002},
}

@article{Majland2020,
  author     = {Majland, Marco and Christensen, Kasper Sangild and Zinner, Nikolaj Thomas},
  title      = {{Quantum thermal transistor in superconducting circuits}},
  journal    = {Physical Review B},
  year       = {2020},
  volume     = {101},
  number     = {18},
  pages      = {184510},
  doi        = {10.1103/physrevb.101.184510},
  url        = {https://doi.org/10.1103/physrevb.101.184510},
}

@article{Rajapaksha2025,
  author     = {Rajapaksha, Anuradhi and Gunapala, Sarath D. and Premaratne, Malin},
  title      = {{Ebers--Moll model inspired equivalent circuit for quantum thermal transistors}},
  journal    = {APL Quantum},
  year       = {2025},
  volume     = {2},
  number     = {2},
  pages      = {026119},
  doi        = {10.1063/5.0270456},
  url        = {https://doi.org/10.1063/5.0270456},
}

@article{Wang2018,
  author     = {Wang, Chen and Chen, Xu-Min and Sun, Ke-Wei and Ren, Jie},
  title      = {{Heat amplification and negative differential thermal conductance in a strongly coupled nonequilibrium spin-boson system}},
  journal    = {Physical Review A},
  year       = {2018},
  volume     = {97},
  number     = {5},
  pages      = {052112},
  doi        = {10.1103/physreva.97.052112},
  url        = {https://doi.org/10.1103/physreva.97.052112},
}

@article{Liu2019,
  author     = {Liu, Huan and Wang, Chen and Wang, Lu-Qing and Ren, Jie},
  title      = {{Strong system-bath coupling induces negative differential thermal conductance and heat amplification in nonequilibrium two-qubit systems}},
  journal    = {Physical Review E},
  year       = {2019},
  volume     = {99},
  number     = {3},
  pages      = {032114},
  doi        = {10.1103/physreve.99.032114},
  url        = {https://doi.org/10.1103/physreve.99.032114},
}

@article{Jiang2015,
  author     = {Jiang, Jian-Hua and Kulkarni, Manas and Segal, Dvira and Imry, Yoseph},
  title      = {{Phonon thermoelectric transistors and rectifiers}},
  journal    = {Physical Review B},
  year       = {2015},
  volume     = {92},
  number     = {4},
  pages      = {045309},
  doi        = {10.1103/physrevb.92.045309},
  url        = {https://doi.org/10.1103/physrevb.92.045309},
}

@article{Wang2019,
  author     = {Wang, Chen and Xu, Dazhi and Liu, Huan and Gao, Xianlong},
  title      = {{Thermal rectification and heat amplification in a nonequilibrium V-type three-level system}},
  journal    = {Physical Review E},
  year       = {2019},
  volume     = {99},
  number     = {4},
  pages      = {042102},
  doi        = {10.1103/physreve.99.042102},
  url        = {https://doi.org/10.1103/physreve.99.042102},
}

@article{Wijesekara2020,
  author     = {Wijesekara, Ravi T. and Gunapala, Sarath D. and Stockman, Mark I. and Premaratne, Malin},
  title      = {{Optically controlled quantum thermal gate}},
  journal    = {Physical Review B},
  year       = {2020},
  volume     = {101},
  number     = {24},
  pages      = {245402},
  doi        = {10.1103/physrevb.101.245402},
  url        = {https://doi.org/10.1103/physrevb.101.245402},
}

@article{Gupt2022,
  author     = {Gupt, Nikhil and Bhattacharyya, Srijan and Das, Bikash and Datta, Subhadeep and Mukherjee, Victor and Ghosh, Arnab},
  title      = {{Floquet quantum thermal transistor}},
  journal    = {Physical Review E},
  year       = {2022},
  volume     = {106},
  number     = {2},
  pages      = {024110},
  doi        = {10.1103/physreve.106.024110},
  url        = {https://doi.org/10.1103/physreve.106.024110},
}

@article{Naseem2020HeatManager,
  author     = {Naseem, M. Tahir and Misra, Avijit and M{\"{u}}stecapl{\i}o{\u{g}}lu, {\"{O}}zg{\"{u}}r E. and Kurizki, Gershon},
  title      = {{Minimal quantum heat manager boosted by bath spectral filtering}},
  journal    = {Physical Review Research},
  year       = {2020},
  volume     = {2},
  number     = {3},
  pages      = {033285},
  doi        = {10.1103/physrevresearch.2.033285},
  url        = {https://doi.org/10.1103/physrevresearch.2.033285},
}

@article{Mandarino2022,
  author     = {Mandarino, Antonio},
  title      = {{Quantum Thermal Amplifiers with Engineered Dissipation}},
  journal    = {Entropy},
  year       = {2022},
  volume     = {24},
  number     = {8},
  pages      = {1031},
  doi        = {10.3390/e24081031},
  url        = {https://doi.org/10.3390/e24081031},
}

@article{Liu2022,
  author     = {Liu, Yu-Qiang and Yu, Deng-Hui and Yu, Chang-Shui},
  title      = {{Common Environmental Effects on Quantum Thermal Transistor}},
  journal    = {Entropy},
  year       = {2022},
  volume     = {24},
  number     = {1},
  pages      = {32},
  doi        = {10.3390/e24010032},
  url        = {https://doi.org/10.3390/e24010032},
}

@article{Ekanayake2024,
  author     = {Ekanayake, Uthpala N. and Gunapala, Sarath D. and Premaratne, Malin},
  title      = {{Improving quantum thermal transistors through feedback-controlled baths}},
  journal    = {APL Quantum},
  year       = {2024},
  volume     = {1},
  number     = {3},
  pages      = {036126},
  doi        = {10.1063/5.0229630},
  url        = {https://doi.org/10.1063/5.0229630},
}

@article{Malavazi2024,
  author     = {Malavazi, Andr{\'{e}} H. A. and Ahmadi, Borhan and Mazurek, Pawe{\l} and Mandarino, Antonio},
  title      = {{Detuning effects for heat-current control in quantum thermal devices}},
  journal    = {Physical Review E},
  year       = {2024},
  volume     = {109},
  number     = {6},
  pages      = {064146},
  doi        = {10.1103/physreve.109.064146},
  url        = {https://doi.org/10.1103/physreve.109.064146},
}

@article{Bhargava2026,
  author     = {Bhargava, Deepika and Chaki, Paranjoy and Bhattacharyya, Aparajita and Sen, Ujjwal},
  title      = {{Quantum transistors for heat flux in and out of working substance parts: harmonic vs transmon and Kerr environs}},
  journal    = {Quantum Science and Technology},
  year       = {2026},
  volume     = {11},
  number     = {4},
  pages      = {045010},
  doi        = {10.1088/2058-9565/ae94a2},
  url        = {https://doi.org/10.1088/2058-9565/ae94a2},
}

@article{Nie2024,
  author     = {Nie, Wenjie and Xu, Junxi and Zhan, Huiya and Chen, Aixi and Lan, Yueheng},
  title      = {{Continuous-variable electromechanical quantum thermal transistors}},
  journal    = {New Journal of Physics},
  year       = {2024},
  volume     = {26},
  number     = {11},
  pages      = {113009},
  doi        = {10.1088/1367-2630/ad8eea},
  url        = {https://doi.org/10.1088/1367-2630/ad8eea},
}

@article{RieraCampeny2019,
  author     = {Riera-Campeny, Andreu and Mehboudi, Mohammad and Pons, Marisa and Sanpera, Anna},
  title      = {Dynamically induced heat rectification in quantum systems},
  journal    = {Physical Review E},
  year       = {2019},
  volume     = {99},
  number     = {3},
  pages      = {032126},
  numpages   = {10},
  month      = mar,
  publisher  = {American Physical Society},
  doi        = {10.1103/PhysRevE.99.032126},
  url        = {https://doi.org/10.1103/PhysRevE.99.032126},
}

@article{Ming2016,
  author     = {Ming, Yi and Li, Hui-Min and Ding, Ze-Jun},
  title      = {{Thermal rectification and negative differential thermal conductance in harmonic chains with nonlinear system-bath coupling}},
  journal    = {Physical Review E},
  year       = {2016},
  volume     = {93},
  number     = {3},
  pages      = {032127},
  doi        = {10.1103/physreve.93.032127},
  url        = {https://doi.org/10.1103/physreve.93.032127},
}

@article{Poyatos1996,
  author     = {Poyatos, J. F. and Cirac, J. I. and Zoller, P.},
  title      = {{Quantum Reservoir Engineering with Laser Cooled Trapped Ions}},
  journal    = {Physical Review Letters},
  year       = {1996},
  volume     = {77},
  number     = {23},
  pages      = {4728--4731},
  doi        = {10.1103/physrevlett.77.4728},
  url        = {https://doi.org/10.1103/physrevlett.77.4728},
}

@article{Verstraete2009,
  author     = {Verstraete, Frank and Wolf, Michael M. and Ignacio Cirac, J.},
  title      = {{Quantum computation and quantum-state engineering driven by dissipation}},
  journal    = {Nature Physics},
  year       = {2009},
  volume     = {5},
  number     = {9},
  pages      = {633--636},
  doi        = {10.1038/nphys1342},
  url        = {https://doi.org/10.1038/nphys1342},
}

@article{Voje2013,
  author     = {Voje, A and Croy, A and Isacsson, A},
  title      = {{Multi-phonon relaxation and generation of quantum states in a nonlinear mechanical oscillator}},
  journal    = {New Journal of Physics},
  year       = {2013},
  volume     = {15},
  number     = {5},
  pages      = {053041},
  doi        = {10.1088/1367-2630/15/5/053041},
  url        = {https://doi.org/10.1088/1367-2630/15/5/053041},
}

@article{Leghtas2015,
  author     = {Leghtas, Z. and Touzard, S. and Pop, I. M. and Kou, A. and Vlastakis, B. and Petrenko, A. and Sliwa, K. M. and Narla, A. and Shankar, S. and Hatridge, M. J. and Reagor, M. and Frunzio, L. and Schoelkopf, R. J. and Mirrahimi, M. and Devoret, M. H.},
  title      = {{Confining the state of light to a quantum manifold by engineered two-photon loss}},
  journal    = {Science},
  year       = {2015},
  volume     = {347},
  number     = {6224},
  pages      = {853--857},
  doi        = {10.1126/science.aaa2085},
  url        = {https://doi.org/10.1126/science.aaa2085},
}

@article{Lescanne2020,
  author     = {Lescanne, Rapha{\"{e}}l and Villiers, Marius and Peronnin, Th{\'{e}}au and Sarlette, Alain and Delbecq, Matthieu and Huard, Benjamin and Kontos, Takis and Mirrahimi, Mazyar and Leghtas, Zaki},
  title      = {{Exponential suppression of bit-flips in a qubit encoded in an oscillator}},
  journal    = {Nature Physics},
  year       = {2020},
  volume     = {16},
  number     = {5},
  pages      = {509--513},
  doi        = {10.1038/s41567-020-0824-x},
  url        = {https://doi.org/10.1038/s41567-020-0824-x},
}

@article{Marquet2024,
  author     = {Marquet, A. and Dupouy, S. and R{\'{e}}glade, U. and Essig, A. and Cohen, J. and Albertinale, E. and Bienfait, A. and Peronnin, T. and Jezouin, S. and Lescanne, R. and Huard, B.},
  title      = {{Harnessing two-photon dissipation for enhanced quantum measurement and control}},
  journal    = {Physical Review Applied},
  year       = {2024},
  volume     = {22},
  number     = {3},
  pages      = {034053},
  doi        = {10.1103/physrevapplied.22.034053},
  url        = {https://doi.org/10.1103/physrevapplied.22.034053},
}

@article{Zhou2022,
  author     = {Zhou, Yan-Hui and Zhang, Xing-Yuan and Liu, Tong and Wu, Qi-Cheng and Shi, Zhi-Cheng and Shen, Hong-Zhi and Yang, Chui-Ping},
  title      = {{Environmentally Induced Photon Blockade via Two-Photon Absorption}},
  journal    = {Physical Review Applied},
  year       = {2022},
  volume     = {18},
  number     = {6},
  pages      = {064009},
  doi        = {10.1103/physrevapplied.18.064009},
  url        = {https://doi.org/10.1103/physrevapplied.18.064009},
}

@article{Nunnenkamp2010,
  author     = {Nunnenkamp, A. and B{\o}rkje, K. and Harris, J. G. E. and Girvin, S. M.},
  title      = {{Cooling and squeezing via quadratic optomechanical coupling}},
  journal    = {Physical Review A},
  year       = {2010},
  volume     = {82},
  number     = {2},
  pages      = {021806},
  doi        = {10.1103/physreva.82.021806},
  url        = {https://doi.org/10.1103/physreva.82.021806},
}

@article{Mirrahimi2014,
  author     = {Mirrahimi, Mazyar and Leghtas, Zaki and Albert, Victor V and Touzard, Steven and Schoelkopf, Robert J and Jiang, Liang and Devoret, Michel H},
  title      = {{Dynamically protected cat-qubits: a new paradigm for universal quantum computation}},
  journal    = {New Journal of Physics},
  year       = {2014},
  volume     = {16},
  number     = {4},
  pages      = {045014},
  doi        = {10.1088/1367-2630/16/4/045014},
  url        = {https://doi.org/10.1088/1367-2630/16/4/045014},
}

@article{Reglade2024,
  author     = {R{\'{e}}glade, U. and Bocquet, A. and Gautier, R. and Cohen, J. and Marquet, A. and Albertinale, E. and Pankratova, N. and Hall{\'{e}}n, M. and Rautschke, F. and Sellem, L.-A. and Rouchon, P. and Sarlette, A. and Mirrahimi, M. and Campagne-Ibarcq, P. and Lescanne, R. and Jezouin, S. and Leghtas, Z.},
  title      = {{Quantum control of a cat qubit with bit-flip times exceeding ten seconds}},
  journal    = {Nature},
  year       = {2024},
  volume     = {629},
  number     = {8013},
  pages      = {778--783},
  doi        = {10.1038/s41586-024-07294-3},
  url        = {https://doi.org/10.1038/s41586-024-07294-3},
}

@article{Marquet2024Autoparametric,
  author     = {Marquet, A. and Essig, A. and Cohen, J. and Cottet, N. and Murani, A. and Albertinale, E. and Dupouy, S. and Bienfait, A. and Peronnin, T. and Jezouin, S. and Lescanne, R. and Huard, B.},
  title      = {{Autoparametric Resonance Extending the Bit-Flip Time of a Cat Qubit up to 0.3 s}},
  journal    = {Physical Review X},
  year       = {2024},
  volume     = {14},
  number     = {2},
  pages      = {021019},
  doi        = {10.1103/physrevx.14.021019},
  url        = {https://doi.org/10.1103/physrevx.14.021019},
}

@article{Putterman2025,
  author     = {Putterman, Harald and Noh, Kyungjoo and Hann, Connor T. and MacCabe, Gregory S. and Aghaeimeibodi, Shahriar and Patel, Rishi N. and Lee, Menyoung and Jones, William M. and Moradinejad, Hesam and Rodriguez, Roberto and Mahuli, Neha and Rose, Jefferson and Owens, John Clai and Levine, Harry and Rosenfeld, Emma and Reinhold, Philip and Moncelsi, Lorenzo and Alcid, Joshua Ari and Alidoust, Nasser and Arrangoiz-Arriola, Patricio and Barnett, James and Bienias, Przemyslaw and Carson, Hugh A. and Chen, Cliff and Chen, Li and Chinkezian, Harutiun and Chisholm, Eric M. and Chou, Ming-Han and Clerk, Aashish and Clifford, Andrew and Cosmic, R. and Curiel, Ana Valdes and Davis, Erik and DeLorenzo, Laura and D'Ewart, J. Mitchell and Diky, Art and D'Souza, Nathan and Dumitrescu, Philipp T. and Eisenmann, Shmuel and Elkhouly, Essam and Evenbly, Glen and Fang, Michael T. and Fang, Yawen and Fling, Matthew J. and Fon, Warren and Garcia, Gabriel and Gorshkov, Alexey V. and Grant, Julia A. and Gray, Mason J. and Grimberg, Sebastian and Grimsmo, Arne L. and Haim, Arbel and Hand, Justin and He, Yuan and Hernandez, Mike and Hover, David and Hung, Jimmy S. C. and Hunt, Matthew and Iverson, Joe and Jarrige, Ignace and Jaskula, Jean-Christophe and Jiang, Liang and Kalaee, Mahmoud and Karabalin, Rassul and Karalekas, Peter J. and Keller, Andrew J. and Khalajhedayati, Amirhossein and Kubica, Aleksander and Lee, Hanho and Leroux, Catherine and Lieu, Simon and Ly, Victor and Madrigal, Keven Villegas and Marcaud, Guillaume and McCabe, Gavin and Miles, Cody and Milsted, Ashley and Minguzzi, Joaquin and Mishra, Anurag and Mukherjee, Biswaroop and Naghiloo, Mahdi and Oblepias, Eric and Ortuno, Gerson and Pagdilao, Jason and Pancotti, Nicola and Panduro, Ashley and Paquette, JP and Park, Minje and Peairs, Gregory A. and Perello, David and Peterson, Eric C. and Ponte, Sophia and Preskill, John and Qiao, Johnson and Refael, Gil and Resnick, Rachel and Retzker, Alex and Reyna, Omar A. and Runyan, Marc and Ryan, Colm A. and Sahmoud, Abdulrahman and Sanchez, Ernesto and Sanil, Rohan and Sankar, Krishanu and Sato, Yuki and Scaffidi, Thomas and Siavoshi, Salome and Sivarajah, Prasahnt and Skogland, Trenton and Su, Chun-Ju and Swenson, Loren J. and Teo, Stephanie M. and Tomada, Astrid and Torlai, Giacomo and Wollack, E. Alex and Ye, Yufeng and Zerrudo, Jessica A. and Zhang, Kailing and Brand{\~{a}}o, Fernando G. S. L. and Matheny, Matthew H. and Painter, Oskar},
  title      = {{Hardware-efficient quantum error correction via concatenated bosonic qubits}},
  journal    = {Nature},
  year       = {2025},
  volume     = {638},
  number     = {8052},
  pages      = {927--934},
  doi        = {10.1038/s41586-025-08642-7},
  url        = {https://doi.org/10.1038/s41586-025-08642-7},
}

@article{PhysRevA.105.012201,
  author     = {Naseem, M. Tahir and M{\"u}stecapl{\i}o{\u{g}}lu, {\"O}zg{\"u}r E.},
  title      = {Antibunching via cooling by heating},
  journal    = {Physical Review A},
  year       = {2022},
  volume     = {105},
  number     = {1},
  pages      = {012201},
  numpages   = {8},
  month      = jan,
  publisher  = {American Physical Society},
  doi        = {10.1103/PhysRevA.105.012201},
  url        = {https://doi.org/10.1103/PhysRevA.105.012201},
}

@article{dxzd-nqrv,
  author     = {Naseem, M. Tahir},
  title      = {Reservoir-engineered mechanical cat states with a driven qubit},
  journal    = {Physical Review A},
  year       = {2026},
  volume     = {113},
  number     = {1},
  pages      = {013732},
  numpages   = {20},
  month      = jan,
  publisher  = {American Physical Society},
  doi        = {10.1103/dxzd-nqrv},
  url        = {https://doi.org/10.1103/dxzd-nqrv},
}

@article{Didier2015,
  author     = {Didier, Nicolas and Bourassa, J{\'{e}}r{\^{o}}me and Blais, Alexandre},
  title      = {{Fast Quantum Nondemolition Readout by Parametric Modulation of Longitudinal Qubit-Oscillator Interaction}},
  journal    = {Physical Review Letters},
  year       = {2015},
  volume     = {115},
  number     = {20},
  pages      = {203601},
  doi        = {10.1103/physrevlett.115.203601},
  url        = {https://doi.org/10.1103/physrevlett.115.203601},
}

@article{Naseem2021ReservoirCooling,
  author     = {Naseem, Muhammad Tahir and M{\"{u}}stecapl{\i}o{\u{g}}lu, {\"{O}}zg{\"{u}}r E.},
  title      = {{Ground-state cooling of mechanical resonators by quantum reservoir engineering}},
  journal    = {Communications Physics},
  year       = {2021},
  volume     = {4},
  number     = {1},
  pages      = {95},
  doi        = {10.1038/s42005-021-00599-z},
  url        = {https://doi.org/10.1038/s42005-021-00599-z},
}

@article{Naseem2026Rectification,
  author     = {Naseem, M Tahir},
  title      = {{Quantum thermal rectification via state-dependent two-photon dissipation}},
  journal    = {Quantum Science and Technology},
  year       = {2026},
  volume     = {11},
  number     = {1},
  pages      = {015011},
  doi        = {10.1088/2058-9565/ae20b7},
  url        = {https://doi.org/10.1088/2058-9565/ae20b7},
}

@article{Naseem_2020,
  author     = {Naseem, M. Tahir and Misra, Avijit and M{\"u}stecapl{\i}o{\u{g}}lu, {\"O}zg{\"u}r E.},
  title      = {Two-body quantum absorption refrigerators with optomechanical-like interactions},
  journal    = {Quantum Science and Technology},
  year       = {2020},
  volume     = {5},
  number     = {3},
  pages      = {035006},
  month      = may,
  publisher  = {IOP Publishing},
  doi        = {10.1088/2058-9565/ab8d89},
  url        = {https://doi.org/10.1088/2058-9565/ab8d89},
}

@article{Touzard2019,
  author  = {Touzard, S. and Kou, A. and Frattini, N. E.
             and Sivak, V. V. and Puri, S. and Grimm, A.
             and Frunzio, L. and Shankar, S. and Devoret, M. H.},
  title   = {{Gated Conditional Displacement Readout of
              Superconducting Qubits}},
  journal = {Physical Review Letters},
  year    = {2019},
  volume  = {122},
  number  = {8},
  pages   = {080502},
  doi     = {10.1103/PhysRevLett.122.080502},
  url     = {https://doi.org/10.1103/PhysRevLett.122.080502}
}

@misc{DataRepository,
  author       = {Naseem, M. Tahir},
  title        = {{Quantum Thermal Transistor via Nonlinear
                   Environmental Coupling}},
  year         = {2026},
  howpublished = {GitHub repository},
  url          = {https://github.com/MuhammadTahirNaseem/Quantum-Thermal-Transistor-via-Nonlinear-Environmental-Coupling},
  note         = {Numerical code for generating the figures}
}

\end{document}